\documentclass[%
 nofootinbib
]{revtex4-2}

\usepackage{graphicx}
\usepackage{dcolumn}
\usepackage{bm}
\usepackage{amsmath}
\usepackage{amssymb}

\usepackage[colorlinks=true,urlcolor=blue,linkcolor=blue,citecolor=blue,linktocpage=true]{hyperref}
\usepackage{color}
\usepackage{comment}
\usepackage{tikz}
\usetikzlibrary{decorations.markings,decorations.pathmorphing}
\usepackage{physics}
\usepackage{dsfont}
\usepackage{color}
\newcommand{\figref}[1]	{{Fig.~\ref{#1}}}
\newcommand{\secref}[1]	{{Sec.~\ref{#1}}}

\renewcommand{\i}{\text{i}}
\newcommand{\bracket}[2]{\langle #1 | #2 \rangle}
\newcommand{\cH}{\mathcal{H}}
\newcommand{\cF}{\mathcal{F}}
\newcommand{\cO}{\mathcal{O}}
\newcommand{\NN}{{\mathbb{N}}}
\newcommand{\ZZ}{{\mathbb{Z}}}
\newcommand{\CC}{{\mathbb{C}}}

\begin{document}

\preprint{APS/123-QED}

\title{From Complexity Geometry to Holographic Spacetime}

\author{Johanna Erdmenger}
    \email{erdmenger@physik.uni-wuerzburg.de}
    \affiliation{Institute for Theoretical Physics and Astrophysics \\ and Würzburg-Dresden Cluster of Excellence ct.qmat, Julius-Maximilians-Universität Würzburg, \\97074 Würzburg, Germany}

\author{Marius Gerbershagen}
    \email{marius.gerbershagen@vub.be}
        \affiliation{Theoretische Natuurkunde, Vrije Universiteit Brussel (VUB) and
    The International Solvay Institutes, Pleinlaan 2, 1050 Brussels, Belgium}

\author{Michal P. Heller}
    \email{michal.p.heller@ugent.be}
    \affiliation{Department of Physics and Astronomy, Ghent University, 9000 Ghent, Belgium}

\author{Anna-Lena Weigel}
    \email{anna-lena.weigel@physik.uni-wuerzburg.de}
    \affiliation{Institute for Theoretical Physics and Astrophysics \\ and Würzburg-Dresden Cluster of Excellence ct.qmat, Julius-Maximilians-Universität Würzburg, \\97074 Würzburg, Germany}
    
\begin{abstract}
\noindent An important conjecture within the AdS/CFT correspondence relates holographic spacetime to the quantum computational complexity of the dual quantum field theory. However, the quantitative understanding of this relation is still an open question. 
In this work, we introduce and study a map between a computational complexity measure and its holographic counterpart from first principles.
We consider quantum circuits built out of conformal transformations in two-dimensional conformal field theory and a complexity measure based on assigning a cost to quantum gates via the Fubini-Study distance.
We find a novel geometric object in three-dimensional anti-de Sitter spacetimes that is dual to this distance. 
This duality also provides a more general map between holographic geometry of anti-de Sitter universes and complexity geometry as defined in information theory, in which each point represents a state and distances between states are measured by the Fubini-Study metric.
We apply the newly found duality to the eternal black hole spacetime and discuss both the origin of linear growth of complexity and the switchback effect within our approach.
\end{abstract}

\maketitle

\section{Introduction}

The question of assigning cost to state preparation in holography~\cite{Maldacena:1997re,Gubser:1998bc,Witten:1998qj} has received significant attention in recent years, building on a conjecture by Susskind relating cost assignment to black hole physics~\cite{Susskind:2014rva}.
The focal notion in this context has been computational complexity, a quantity from quantum information counting how many computation steps are necessary to prepare a certain target state from a fixed reference state~\cite{Aaronson:2016vto,Chapman:2021jbh}.
In chaotic systems of finite size, computational complexity is expected to show a number of  features that are universal,  in the sense that they hold for any chaotic system and any reasonable definition of complexity. These expected universal features  were conjectured to be related to geometric features of anti-de Sitter (AdS) black hole geometries in \cite{Susskind:2014rva}. They involve the growth of the size of black hole interiors with time.
In line with the universality expectation, these features are probed by many gravity observables.
The most prominently studied holographic complexity proposals are the so-called ``complexity=volume'' \cite{Susskind:2014rva,Stanford:2014jda}, ``complexity=action'' \cite{Brown:2015bva,Brown:2015lvg} and ``complexity=volume 2.0''~\cite{Couch:2016exn} proposals.  Recently, an infinite class of such gravitational observables was put forward under the slogan ``complexity=anything''  \cite{Belin:2021bga,Belin:2022xmt}.

Understanding holographic complexity in terms of a dual microscopic description triggered progress on defining and studying computational complexity within quantum field theory (QFTs) (see~\cite{Chapman:2021jbh} for a review). However, up to now the relation to holographic complexity has almost always been restricted to qualitative comparisons, due to to the fact that soluble examples are either free QFTs or models lacking control over either the boundary or the bulk side.
Here, for the first time, we propose a map between computational complexity and a geometric object in the gravity theory which is derived from first principles and which can be naturally applied to the most relevant case -- AdS black hole spacetimes and dual thermofield double (TFD) states~\cite{Maldacena:2001kr}.

Much of the previous work on computational complexity in QFTs is based on the  framework proposed by Nielsen in~\cite{Nielsen} to bound complexity of discrete circuits using differential geometry tools.
Instead of counting discrete gates that belong to the native quantum computing language, in the Nielsen approach the circuit evolution proceeds in a continuous manner by applying a path ordered exponential
\begin{equation}
\label{eq.circuit}
  \ket{\psi(\tau)} = \mathcal{P} \exp\left(i \int_0^\tau d\tau' Q(\tau')\right)\ket{\psi(0)}.
\end{equation}
Different circuits performing the same task are distinguished by a cost function $F[\ket{\psi(\tau)},Q(\tau)]$ that measures how expensive the application of $Q(\tau)$ onto the state $\ket{\psi(\tau)}$ is. Such a cost function can viewed as weighting components of $Q(\tau)$ in a basis of generators of infinitesimal gates viewed as ultimate circuit building blocks. The complexity is then defined as the minimum of the total cost,
\begin{equation}
\label{eq.complexitydef}
  C = \min \int_0^{\tau_f} d\tau\, F[\ket{\psi(\tau)},Q(\tau)] \, ,
\end{equation}
subject to the condition that they connect fixed reference $\ket{\psi(0)}$ and target $\ket{\psi(\tau_f)}$ states.
A cost function $F$ defines a {\it Finsler geometry}~\cite{Nielsen} if it is is smooth, positive, positively homogeneous of degree one in its second argument and obeys the triangle inequality. In this case, the target and reference states are represented by manifold points and the complexity by the length of the shortest path between them, i.e.~by the geodesic.
The following question then arises  naturally: 
\begin{quote}
How is this auxiliary complexity geometry for properly understood circuits in holographic QFTs encoded in the gravity dual description?
\end{quote}
This important question motivates the present work. There are three key conceptual problems that we have to address in view of answering this question: What is meant  by $\tau$ in holography? What constitutes a gate set in a QFT? What QFT cost functions can be interpreted holographically? In our proposal we are guided by the observation that the only interface to translate between the boundary and the bulk comes from the identification of source terms in path integrals in holographic QFTs with asymptotic boundary conditions for bulk fields~\cite{Gubser:1998bc,Witten:1998qj,Skenderis:2008dh,Skenderis:2008dg}. Therefore, it is natural to identify the circuit parameter~$\tau$ with the physical time $t$ at the boundary~\cite{Erdmenger:2021wzc}
\begin{equation}
\tau \equiv t.
\end{equation}
As a consequence, the circuit generator $Q(\tau)$ has to be identified with the physical Hamiltonian~$H(t)$,
\begin{equation}
Q(\tau) \equiv H(t).
\end{equation}
Different QFT source configurations correspond to controllable modifications of the Hamiltonian obtained by adding local primary operators. Therefore, these operators constitute natural generators of the infinitesimal gates. Moreover, constant $t$ slices naturally define states.  Considering such slices therefore naturally introduces  reference and target states $|\psi(0)\rangle$ and $|\psi(\tau_f)\rangle$.
This perspective on obtaining gravity duals for a quantum circuits was outlined in our previous work \cite{Erdmenger:2021wzc}.
Here, we take the crucial further step of connecting the gravity duals of quantum circuits with notions of computational complexity.

\begin{figure}
    \centering
    \begin{tikzpicture}
    \begin{scope}[scale=0.9]
        \shade[ball color = gray!40, opacity = 0.4] (0,0) circle (2cm);
        \draw (0,0) circle (2cm);
        \draw (-2,0) arc (180:360:2 and 0.6);
        \draw[dashed] (2,0) arc (0:180:2 and 0.6);
        \fill[fill=red!50!blue] (0,-2) circle (1.5pt);
        \draw (0,-2) node[below] {\small $\ket{\psi(t_1)}$};
        \fill[fill=black] (0,2) circle (1.5pt);
        \draw (0,2) node[above] {\small $\ket\chi$};
        \fill[fill=red!50!blue] ({{cos(290)*2}},0) circle (1.5pt);
        \draw ({{cos(290)*2}},0) node[right] {\small $\ket{\psi(t_2)}$};
        \draw[color=red!50!blue] (0,-2) to[out=35,in=285] ({{cos(290)*2}},0);
        \draw[dashed,thin] (0,-2) -- (0,-0.2) -- ({{cos(290)*2}},0);
        \draw[thin] (0,-0.9) to[out=0,in=280] ({{cos(290)*2*0.6}},{{-0.6*0.2}});
        \draw[thin] (0.2,-0.4) node {\small $2\theta$};
    \end{scope}
    \begin{scope}[scale=0.6,shift={(10,0)}]
      \path[left color=black!10,right color=black!2] (-2,3) arc (180:0:2 and 0.75) -- (2,-3) arc(0:180:2 and 0.75) -- (-2,3);
      \draw (0,-3) ellipse (2 and 0.75);
      \draw (0,3) ellipse (2 and 0.75);
      \path[left color=blue!40!white,right color=white,opacity=0.7] (-2,1) arc (180:0:2 and 0.75) -- (2,-1) arc(0:180:2 and 0.75) -- (-2,1);
      \path[left color=blue!40!white,right color=blue!10!white,opacity=0.7] (-2,1) arc (180:360:2 and 0.75) -- (2,-1) arc(360:180:2 and 0.75) -- (-2,1);
      \draw[color=blue!80!black,thick] (0,-1) ellipse (2 and 0.75);
      \draw (2,-1) node[right] {$\ket{\psi(0}$};
      \draw[color=blue!80!black,thick] (0,1) ellipse (2 and 0.75);
      \draw (2,1) node[right] {$\ket{\psi(t_f)}$};
      \draw[->] (2.3,-3) -- node[midway,right] {$t=\tau$} (2.3,-1.95);
      \draw (-2,-3) -- (-2,3);
      \draw (2,-3) -- (2,3);
      \path[left color=black!40,right color=black!5,opacity=0.3] (-2,3) arc (180:360:2 and 0.75) -- (2,-3) arc(360:180:2 and 0.75) -- (-2,3);
    \end{scope}
    \end{tikzpicture}
    \caption{Illustration of the map between a distance measure in the complexity geometry on the left and a geometric object in the asymptotically AdS spacetime on the right. Any two states $\ket{\psi(t_1)},\ket{\psi(t_2)}$ can be put on a Bloch sphere spanned by $\ket{\psi(t_1)}$ and an orthogonal state $\ket{\chi}$ obtained by subtracting from $\ket{\psi(t_2)}$ the part parallel to $\ket{\psi(t_1)}$, i.e.~$\ket{\psi(t_2)} \propto \ket{\chi} + \bracket{\psi(t_1)}{\psi(t_2)}\ket{\psi(t_1)}$. The geodesic distance for the Fubini-Study metric between $\ket{\psi(t_1)}$ and $\ket{\psi(t_2)}$ is then the angle $\theta$ on this Bloch sphere.  These two states live on different time slices at the boundary of the same AdS geometry shown on the 
    right-hand side. The infinitesimal distance in the complexity geometry on the left between the states $\ket{\psi(t_1=t)}$ and $\ket{\psi(t_2=t+dt)}$ manifests itself as a geometric object in the AdS space on the right. Therefore, the total cost also acquires a geometric dual localized in between the two time slices $t=0$ and $t=t_f$ in the AdS space. For optimal circuits, this geometric object becomes a gravity dual to the complexity.}
    \label{fig:complexity-geometry-vs-AdS-geometry}
\end{figure}
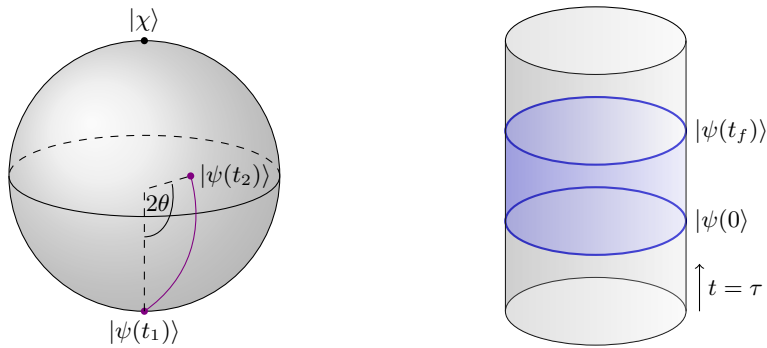

We consider a computational complexity definition based on the Fubini-Study distance as cost function. The Fubini-Study distance is a natural distance measure in Hilbert space.
For two infinitesimally separated states $\ket\psi$ and~$\ket\psi+\ket{\delta\psi}$,  it takes the form
\begin{equation}
  ds^2_\text{FS} = \frac{\bracket{\delta\psi}{\delta\psi}}{\bracket\psi\psi} - \frac{\bracket{\delta\psi}{\psi}\bracket{\psi}{\delta\psi}}{\bracket\psi\psi^2}.
  \label{eq:Fubini-Study-line-element}
\end{equation}
The Fubini-Study distance is the unique Riemannian metric on projective Hilbert space\footnote{The projective Hilbert space arises from identifying $\ket\psi \sim \alpha \ket\psi$ for $\alpha \in \CC$, i.e.~it is the space of properly normalized physically indistinguishable Hilbert space elements.} invariant under unitary transformations \cite{bengtsson_zyczkowski_2017}.
The geodesic distance $\theta$ between two states $\ket{\psi(t_1)}$ and $\ket{\psi(t_2)})$ for this metric is given by (see \figref{fig:complexity-geometry-vs-AdS-geometry})
\begin{equation}
    \cos(\theta)^2 = \frac{\bracket{\psi(t_1)}{\psi(t_2)}\bracket{\psi(t_2)}{\psi(t_1)}}{\bracket{\psi(t_1)}{\psi(t_1)}\bracket{\psi(t_2)}{\psi(t_2)}},
\end{equation}
which implies that $\theta$ is bounded from above by $\pi/2$.

How can this ansatz define a non-trivial complexity measure?
The answer to this question relies on the fact that due to restricting to Hamiltonians obtained by turning on specific sources, we cannot move along some of the directions in the Hilbert space and therefore the shortest path is not necessarily a geodesic.
In other words, by only allowing directions corresponding to a subset of source deformations in the QFT, we assign infinite cost to the other Hilbert space directions.
Moreover, we define the cost function to be the square of the Fubini-Study line element,
\begin{equation}
    F_\text{FS} = ds^2_\text{FS}.
\end{equation}
This is a slight departure from the formalism of \cite{Nielsen}, as it implies that the cost function is not positively homogeneous of degree one and hence does not define a Finsler metric in our setup.
However, similar notions of complexity associated to cost functions which are positively homogeneous of degree greater than one were introduced into the computational complexity setup for QFTs in \cite{Jefferson:2017sdb} with the justification that it provides a better match to the ``complexity=volume'' proposal\footnote{In a restricted special case this also applies to the Fubini-Study complexity: for states that are perturbative conformal transformations of the vacuum, the Fubini-Study complexity we are considering here was found to be proportional to a difference in ``complexity=volume'' between the vacuum and the target state up to the third order in perturbation theory \cite{Flory:2020dja,Flory:2020eot,Erdmenger:2021wzc}. This agreement would not hold if one were to use the Fubini-Study line element instead of its square in the definition.} (see also \cite{Hackl:2018ptj,Chapman:2018hou,Bueno:2019ajd} for further discussion on the merits and drawbacks of this choice).
For us, choosing the cost function as the square of the Fubini-Study line element simplifies the dual bulk description while also allowing for a complexity measure that fulfills important properties postulated to hold for complexity measures in the AdS/CFT setting, as we will see in the following.

A key insight of our present work comes from recognizing that the holographic dictionary naturally provides information about overlaps of states and, related to them, correlation functions.
The Fubini-Study cost function utilizes precisely this information, as it reduces to the variance of $H(t)$ when applied to one layer of the circuit~\eqref{eq.circuit},
\begin{equation}
      F_\text{FS} = \bra{\psi(t)}H(t)^2\ket{\psi(t)} - \bra{\psi(t)}H(t)\ket{\psi(t)}^2.
      \label{eq:Fubini-Study-cost-function}
\end{equation}
This quantity was previously investigated in the computational complexity context in \cite{Chapman:2017rqy,Caputa:2018kdj,Flory:2020eot,Flory:2020dja,Erdmenger:2021wzc}. As $H(t)$ is a sum over integrals of local operators, the Fubini-Study cost~\eqref{eq:Fubini-Study-cost-function} is a linear combination of two-point functions in a state $|\psi(t)\rangle$. Since there is a systematic holographic procedure for calculating two-point functions in any geometric state, the Fubini-Study metric~\eqref{eq:Fubini-Study-cost-function} is not only a natural choice from the computational complexity point of view, but also has a natural holographic realization. However, this bulk realization is in general nontrivial: as the states of interest are by construction time-dependent, it requires knowledge of non-equilibrium two-point functions associated to an evolving gravitational background. The dynamics of such two-point functions can be studied, as was done for instance in~\cite{Chesler:2011ds,Chesler:2012zk,Keranen:2014lna}, but in general is accessible only via means of numerical holography~\cite{Liu:2018crr}. This makes the computation of \eqref{eq:Fubini-Study-cost-function} rather challenging to pursue in a generic setup.

In our work we strive for more, namely for obtaining access to such two-point functions just from the knowledge of the geometry itself\footnote{Obtaining observables just from the geometry itself is possible for one-point functions of local operators~\cite{Witten:1998qj,Gubser:1998bc,Bianchi:2001kw,Skenderis:2008dh,Skenderis:2008dg}, Wilson loops~\cite{Maldacena:1998im,Rey:1998ik} and the fine-grained entropy~\cite{Ryu:2006bv,Hubeny:2007xt,Lewkowycz:2013nqa,Dong:2016hjy}. In contrast, the computation of higher R{\'e}nyi entropies requires backreaction \cite{Dong:2016fnf} and hence these entropies are not encoded in the non-backreacted geometry itself.}. This turns out to be possible for circuits holographically represented by the pure gravity sector of AdS$_3$ holography. Such circuits are built from insertions of the energy-momentum tensor in two-dimensional CFT  acting on an energy eigenstate or a TFD state. Alternatively, we may view such circuits as realizing a gradual change in the state via conformal transformations. 
See~\cite{Magan:2018nmu,Caputa:2018kdj,Akal:2019hxa,Erdmenger:2020sup,Flory:2020eot,Flory:2020dja} and~\cite{Belin:2018bpg,Flory:2018akz,Flory:2019kah,Chagnet:2021uvi,Caputa:2022zsr} for earlier works on these circuits on, respectively the field theory and gravity sides of the duality. In particular, the authors of \cite{Chagnet:2021uvi} studied Fubini-Study cost for circuits implementing global conformal transformations in $d\geq 2$ and derived a bulk dual to the Fubini-Study metric for this particular class of circuits.

 Building on previous work \cite{Erdmenger:2021wzc}, in the present paper we map the Fubini-Study distance \eqref{eq:Fubini-Study-cost-function} along the boundary circuit to a geometric quantity in the gravity theory using the AdS/CFT dictionary. From a broader perspective, we thus provide a precise relation between infinitely dimensional complexity geometry associated with the Fubini-Study cost studied in~\cite{Flory:2020eot,Flory:2020dja} and the holographic geometry of AdS universes. The gravity expression we find applies to all asymptotically AdS$_3$ geometries without matter fields dual to conformal transformations of the vacuum state and excited as well as thermal states.
This result opens up the possibility for deriving further gravity duals of quantum information quantities. It can be also thought of as a natural, yet crucial in the context of holographic complexity generalization of the previous AdS/CFT studies of distance measures between states such as the Fisher information metric \cite{Miyaji:2015woj,Lashkari:2015hha,Bak:2015jxd,Trivella:2016brw,Miyaji:2016fse,Banerjee:2017qti,Flory:2017ftd,Bak:2017rpp,Belin:2018bpg,Xu:2018ngh,Chen:2018vkw,Dimov:2020fzi,Tsuchiya:2021hvg}.

Furthermore, we also investigate which features expected from computational complexity in finite size chaotic systems are reproduced by the Fubini-Study complexity measure that we consider.
In the perhaps most interesting case of the time-evolved TFD state, we find a linear growth of complexity, matching the expectations from \cite{Susskind:2014rva}.
To our knowledge, this is the first time that this feature is found in a scenario where both the bulk and boundary sides are under control and the equality of both descriptions is derived from the AdS/CFT dictionary.
Moreover, we apply the gravity expression -- that we found to be dual to the Fubini-Study distance in bulk geometries without matter fields -- to a class of shockwave geometries sourced by sharply concentrated bulk matter.
This reproduces a characteristic time delay in the growth of complexity known as the switchback effect \cite{Stanford:2014jda,Susskind:2014jwa}, a further important feature of complexity in finite size systems. 
However for the states dual to the shockwave geometries, the Fubini-Study distance \eqref{eq:Fubini-Study-cost-function}  is by definition constant in time  and hence cannot  show any sign of the switchback effect. 
Therefore, in this case our new geometric bulk quantity is no longer dual to the Fubini-Study distance.

In summary, we have constructed a gravity observable that satisfies the main features required for a holographic complexity measure -- linear growth and the switchback effect -- whose field theory dual is known for all geometries without matter fields through a first principles derivation.

The outline of our paper is as follows.
We start in \secref{sec:gravity-dual-to-circuit} with a review of the construction from \cite{Erdmenger:2021wzc} of the bulk dual to a circuit of conformally transformed CFT states.
In \secref{sec:bulk-dual-Fubini-Study}, we construct the bulk dual to the Fubini-Study distance.
The Fubini-Study complexity of the time-evolved TFD state is studied in \secref{sec:complexity} while \secref{sec:switchback} deals with the question whether it can reproduce the switchback effect.
We close in Secs.~\ref{sec:discussion} and \ref{sec:outlook} with a discussion and outlook. In App.~\ref{sec:SL(2,R)-circuits}, we comment on the differences between our work and \cite{Chagnet:2021uvi}. App.~\ref{sec:further-shockwaves} corroborates our results on the switchback effect discussed in \secref{sec:switchback} by studying it in different shockwave geometries.

\section{Setup: Holographic dual to quantum circuits}
\label{sec:gravity-dual-to-circuit}
As outlined in the introduction, studying holographically circuits generated by arbitrarily smeared local operators in holographic QFTs is of key importance for providing a microscopic understanding of holography of complexity. The simplest available setting is the energy-momentum sector of two-dimensional holographic conformal field theories, which generate (in general, local) conformal transformations. In \cite{Erdmenger:2021wzc}, we presented a general prescription for constructing an exact gravity dual for quantum circuits generated by conformal transformations. Here, we briefly review these results, as they allow us to study explicitly the gravity dual to the Fubini-Study cost and, in the eternal black hole geometry, also to the associated complexity.

We consider a two-dimensional CFT in the standard Euclidean framework obtained by analytic continuation $t \to it$ from Lorentzian signature.
In two dimensions, conformal transformations generate two copies of the Virasoro group whose group elements are orientation preserving diffeomorphisms
\begin{equation}
    z \rightarrow f(z),
\end{equation}
of the complex coordinate $z=t+i\phi$ where $t$ is the time coordinate and $\phi \in [0,2\pi)$ the angular coordinate.
In this setup, the quantum circuit we consider corresponds to a path $f(\tau, z)$ through the space of diffeomorphisms.
As a function of the parameter $\tau$, $f(\tau,z)$ determines the conformal transformation that when applied onto the reference state $\ket{\psi(0)}$ yields the state $\ket{\psi(\tau)}$.
Infinitesimal changes $f(\tau,z) \to f(\tau+d\tau,z)$ along the path are generated by~\cite{Magan:2018nmu,Caputa:2018kdj},
\begin{equation}
    Q(\tau) = \int_{0}^{2\pi} \frac{d\phi}{2\pi} \epsilon(\tau,z) T(z) = \sum_{n=-\infty}^{\infty}\epsilon_{-n}(\tau)L_n.
    \label{eq:circuit-Hamiltonian}
\end{equation}
Here, $T(z)=\sum_n L_n e^{n z}$ is the energy-momentum tensor with Virasoro generators $L_n$ obeying the venerable Virasoro algebra
\begin{equation}
  [L_n,L_m] = (n-m)L_{n+m} + \frac{c}{12}(n^3-n)\delta_{n+m,0},
\end{equation}
and $\epsilon(\tau,z)$ admits the Fourier expansion $\epsilon(\tau,z) = \sum_n \epsilon_n(\tau)e^{n z}$.
The infinitesimal diffeomorphism $z \to z + \epsilon(\tau,z)$ is related to the path $f(\tau,z)$ by the multiplication law of the Virasoro group, giving
\begin{equation}
    \epsilon(\tau, f(\tau, z))=\frac{d}{d \tau} f(\tau, z).
    \label{eq:epsilon-coefficient}
\end{equation}

There are two ways to interpret this circuit gravitationally depending on whether the variable $\tau$ is taken to be an external auxiliary parameter or identified with the physical time.
In the former case, each value of $\tau$ is associated to a different bulk geometry where for all $\tau$, the state $\ket{\psi(\tau)}$ lives on the same time slice in physical time (say at $t = 0$) on the AdS boundary.
For conformal transformations acting on pure states of the CFT, these bulk geometries are Ba{\~n}ados geometries.
On the other hand, if the parameter $\tau$ is identical to the physical time $t$, the states $\ket{\psi(\tau)}$ live on different time slices on the boundary of the same asymptotically AdS spacetime (see \figref{fig:complexity-geometry-vs-AdS-geometry}).
In this case, the non-trivial time evolution is generated by turning on a source term for the energy-momentum tensor in the path integral picture.
It is the latter picture that we will use in the following.

Let us now describe in more detail how to obtain the correct source term for a given circuit determined by the path $f(t,z)$.
As the energy-momentum tensor on the boundary is sourced by the boundary metric $g^{(0)}_{ij}$, we need to consider the CFT on a non-trivial background metric.
This background metric is non-trivial but fixed and thus this procedure does not lead to the introduction of dynamical gravity on the boundary.
The exact form of the background metric is determined by demanding that the physical Hamiltonian $H(t)$ and the sum of the left- and right-moving conformal transformation generators $Q(t)$ and $\bar{Q}(t)$ are equal,
\begin{equation}
    H(t)=\int d\phi \sqrt{\det(g^{(0)})}T^{t}_{\phantom{t}t} \stackrel{!}{=} Q(t) + \bar{Q}(t).
    \label{eq:Hamiltonian_identification}
\end{equation}
This condition ensures that the sequence of states generated by time-evolution in the background $g^{(0)}_{ij}$ is the same as the sequence of states generated by the path ordered exponential \eqref{eq.circuit} together with its right-moving counterpart.

The solution of \eqref{eq:Hamiltonian_identification} is given as follows.
At $t<0$, before the circuit begins acting, the background metric for the CFT is given by $ds^2_{(0)}=dzd\bar{z}$ and the Hamiltonian is the standard CFT Hamiltonian $H(t)=L_0+\bar{L}_0$.
For $0 \leq t \leq t_f$, the circuit implements non-trivial conformal transformations.
In this time range, the boundary metric is given by
\begin{equation}
  ds^2_{(0)} = \left(\frac{\partial f}{\partial z}\right)\left(\frac{\partial \bar{f}}{\partial z}\right)dz^2 + \left(\left(\frac{\partial f}{\partial z}\right)\left(\frac{\partial \bar{f}}{\partial \bar{z}}\right) + \left(\frac{\partial f}{\partial \bar{z}}\right)\left(\frac{\partial \bar{f}}{\partial z}\right)\right)dzd\bar{z} + \left(\frac{\partial f}{\partial \bar{z}}\right)\left(\frac{\partial \bar{f}}{\partial \bar{z}}\right)d\bar{z}^2, \quad 0 \leq t \leq t_f
  \label{eq:boundary-metric}
\end{equation}
where the derivatives acting on $f \equiv f(t,z)$ and $\bar{f} \equiv \bar{f}(t,\bar{z})$ in this expression are all non-vanishing because $t = (z+\bar{z})/2$ depends implicitly on $z$ and $\bar{z}$.
Note that the metric in \eqref{eq:boundary-metric} is flat.
This property, which is special to two dimensions, means that the non-trivial time-evolution we are after is obtained simply by deforming the timelike slices on which the states are defined \footnote{The fact that we can restrict to flat metrics can be justified as follows. An arbitrary curved metric in two dimensions can be written as a Weyl transformation $e^{-2\omega}ds^2_{(0)}$ of a flat metric. Under this transformation, the energy-momentum tensor acquires an additional term which is constant, i.e.~proportional to the identity operator
  \begin{equation}
    T_{i j} \rightarrow T_{i j}+\frac{c}{6}\left(\partial_i \omega \partial_j \omega-\frac{1}{2} g^{(0)}_{i j} \partial^k \omega \partial_k \omega-\nabla^{(0)}_i \nabla^{(0)}_j \omega+g^{(0)}_{i j} \nabla_{(0)}^k \nabla^{(0)}_k \omega\right).
  \end{equation}
  Therefore, also the Hamiltonian $H(t)$ changes by a constant term under this transformation.
  However, the circuit Hamiltonians $Q(t)$ and $\bar{Q}(t)$ do not include any terms proportional to the identity operator, thereby allowing us to exclude curved background metrics.
}.
Finally, at time $t>t_f$ we have arrived at the target state.
In this range of the time coordinate the function $f(t,z)=f(t_f,z)$ is independent of $t$ and the metric \eqref{eq:boundary-metric} reduces to \footnote{It is possible to transform back to a frame in which the metric is $ds^2_{(0)}=dzd\bar{z}$ by applying a residual flatness preserving Weyl transformation
  \begin{equation}
    d s^2_{(0)} \rightarrow e^{-2 \omega} d s^2_{(0)} = \frac{1}{f'(t_f,F(t_f,f(t,z)))} d s^2_{(0)},
  \end{equation}
  where $f'(t,z)$ denotes the derivative w.r.t.~the second argument.
  Since this transformation leaves the final state invariant up to an overall phase factor, it is not strictly necessary and we will omit it in the following.}
\begin{equation}
    ds^2_{(0)} = \left(\frac{\partial f(t_f,z)}{\partial z}\right)\left(\frac{\partial \bar{f}(t_f,\bar{z})}{\partial \bar{z}}\right)dzd\bar{z}, \quad t>t_f.
\end{equation}

Finally, in order to obtain the bulk metric we employ the Fefferman-Graham expansion \cite{Fefferman:1985, deHaro:2000vlm},
\begin{equation}
  d s^2=\frac{d r^2}{r^2}+\left(\frac{1}{r^2} g_{i j}^{(0)}+g_{i j}^{(2)}+r^2 g_{i j}^{(4)}\right) d x^i d x^j,
  \label{eq:Fefferman-Graham-expansion}
\end{equation}
where 
\begin{equation}
  g_{i j}^{(2)}=-\frac{1}{2} R^{(0)} g_{i j}^{(0)}-\frac{6}{c}\left\langle T_{i j}\right\rangle \quad \text { and } \quad g_{i j}^{(4)}=\frac{1}{4}\left(g^{(2)}\left(g^{(0)}\right)^{-1} g^{(2)}\right)_{i j}.
  \label{eq:coefficients-Fefferman-Graham-expansion}
\end{equation}
Unlike in higher dimensions, the Fefferman-Graham expansion truncates and therefore the expression \eqref{eq:Fefferman-Graham-expansion} is valid for all $r$.
The energy-momentum tensor expectation values in \eqref{eq:coefficients-Fefferman-Graham-expansion} are determined from the expectation values $\left\langle T_{i j}\right\rangle$ in the background $ds^2_{(0)}=dz d\bar{z}$ by the same coordinate transformation $z \to f(t,z), \bar{z} \to \bar{f}(t,\bar{z})$ that leads to the expression~\eqref{eq:boundary-metric} for the boundary metric.
For example, for a sequence of states $\ket{\psi(t)}$ given by conformal transformations of the vacuum state, the expectation values in the background $ds^2_{(0)}=dzd\bar{z}$ are given by $\left\langle T_{zz}\right\rangle = \left\langle T_{\bar{z}\bar{z}}\right\rangle = -c/24$, $\left\langle T_{z\bar{z}}\right\rangle = 0$ and transform to
\begin{equation}
    \begin{aligned}
    \left\langle T_{zz}\right\rangle &= -\frac{c}{24}\left(\left(\frac{\partial f}{\partial z}\right)^2 + \left(\frac{\partial \bar{f}}{\partial z}\right)^2\right), \quad \left\langle T_{\bar{z}\bar{z}}\right\rangle = -\frac{c}{24}\left(\left(\frac{\partial f}{\partial \bar{z}}\right)^2 + \left(\frac{\partial \bar{f}}{\partial \bar{z}}\right)^2\right),\\
    \left\langle T_{z\bar{z}}\right\rangle &= -\frac{c}{24}\left(\left(\frac{\partial f}{\partial z}\right)\left(\frac{\partial f}{\partial \bar{z}}\right) + \left(\frac{\partial \bar{f}}{\partial z}\right)\left(\frac{\partial \bar{f}}{\partial \bar{z}}\right)\right)
    \end{aligned}
\end{equation}
in the background \eqref{eq:boundary-metric}.
This procedure yields a bulk dual to the quantum circuit generated by $Q(t)$ and $\bar{Q}(t)$, in which the entire circuit is encoded in the evolution of a single bulk geometry. 

\section{Holographic dual to Fubini-Study distance}
\label{sec:bulk-dual-Fubini-Study}

The goal of this section is to derive the gravitational dual to the Fubini-Study distance \eqref{eq:Fubini-Study-line-element} in the circuit construction described in \secref{sec:gravity-dual-to-circuit}. From a high level point of view, this crucial step provides a bridge between an auxiliary circuit geometry defined in terms of the boundary quantities and the bulk geometry of AdS.

\begin{figure}
    \centering
  \begin{tikzpicture}[scale=0.9]
    \def\i{0}
    \def\radius{2}
    \def\N{3}
    \def\betatilde{60}
    \def\beta{\betatilde/\N}
    \def\eradiusx{\radius}
    \def\eradiusy{\radius/2.5}
    \def\deltay{2.1}
    \begin{scope}
      \path[left color=black!10,right color=black!2] (-\eradiusx,\deltay) arc (180:0:{{\eradiusx}} and {{\eradiusy}}) -- (\eradiusx,-\deltay) arc(0:180:{{\eradiusx}} and {{\eradiusy}}) -- (-\eradiusx,\deltay);
      \draw (0,{{-\deltay}}) ellipse ({{\eradiusx}} and {{\eradiusy}});
      \draw (0,{{\deltay}}) ellipse ({{\eradiusx}} and {{\eradiusy}});
      \draw ({{-\eradiusx}},{{-\deltay}}) -- ({{-\eradiusx}},{{\deltay}});
      \draw ({{\eradiusx}},{{-\deltay}}) -- ({{\eradiusx}},{{\deltay}});
      \draw (0,0) ellipse ({{\eradiusx}} and {{\eradiusy}});
      \draw[color=red!80!black] ({{\eradiusx*sin(\betatilde)}},{{-\eradiusy*cos(\betatilde)}}) node[below,inner sep=1pt,color=black] {$\phi_1$} to[out={{110+\betatilde}},in={{-80}}] node[midway,left,inner sep=6pt] {$\ell$} (0,{{\eradiusy}}) node[above,inner sep=1pt,color=black] {$\phi_2$};
      \path[left color=black!40,right color=black!5,opacity=0.3] (-\eradiusx,\deltay) arc (180:360:{{\eradiusx}} and {{\eradiusy}}) -- (\eradiusx,-\deltay) arc(360:180:{{\eradiusx}} and {{\eradiusy}}) -- (-\eradiusx,\deltay);
      \draw[->] (0,{{-\deltay-(\eradiusy*1.25)}}) to[out=0,in=-150] node[midway,below] {\small{$\phi$}} ({{(\eradiusx*1.25)*sin(60)}},{{-\deltay-(\eradiusy*1.25)*cos(60)}});
      \draw[->] (2.4,-2) -- node[midway,right] {$t_1=t_2=t$} (2.4,-0.95);
    \end{scope}
    \begin{scope}[shift={(9,0)}]
      \path[left color=black!10,right color=black!2] (-\eradiusx,\deltay) arc (180:0:{{\eradiusx}} and {{\eradiusy}}) -- (\eradiusx,-\deltay) arc(0:180:{{\eradiusx}} and {{\eradiusy}}) -- (-\eradiusx,\deltay);
      \draw[thin] ({{-\eradiusx}},{{-\deltay}}) -- ({{-\eradiusx}},{{\deltay}});
      \draw[thin] ({{+\eradiusx}},{{-\deltay}}) -- ({{+\eradiusx}},{{\deltay}});
      \foreach \k in {{{-\deltay}},0,{{\deltay}}}
      {
        \draw[thin] (0,{{\k}}) ellipse ({{\eradiusx}} and {{\eradiusy}});
        \fill[black] (0,{{\k}}) circle (0.05cm);
      }
      \def\alpha{min(\delta,180-\delta)}
      \def\signflip{sign(90-\delta)}
      \def\r{atan(sqrt(abs((1+tan(\phi)*tan(\phi))/(tan(\alpha)*tan(\alpha)+1e-2-tan(\phi)*tan(\phi)))))}
      \foreach \j/\color [evaluate=\j as \jminusone using {int(\j-1)}] in {1/red!80!black,2/blue!80!black,3/green!80!black}
      {
        \def\delta{(\beta+180/\N*(\j-1))}
        \draw[color=\color] plot[domain=-\alpha:\alpha,smooth,variable=\phi] ({\r*\eradiusx/90*sin((\phi+\signflip*\alpha)*\N)},{\r*\eradiusy/90*cos((\phi+\signflip*\alpha)*\N)});
      }
      \draw[color=green!80!black] ({{cos(180+\betatilde/2)*\eradiusx*0.78-0.13}},{{sin(180+\betatilde/2)*\eradiusy*0.78+0.41}}) node {$\ell_2$};
      \draw[color=blue!80!black] ({{cos(90+\betatilde/2)*\eradiusx*0.35}},{{sin(90+\betatilde/2)*\eradiusy*0.35}}) node {$\ell_1$};
      \draw[color=red!80!black] ({{cos(\betatilde/2)*\eradiusx*0.53-0.01}},{{sin(\betatilde/2)*\eradiusy*0.53}}) node {$\ell_0$};
      \draw  ({{\eradiusx*sin(\betatilde)}},{{-\eradiusy*cos(\betatilde)}}) node[below,inner sep=1pt,color=black] {$\phi_1$};
      \draw (0,{{\eradiusy}}) node[above,inner sep=1pt,color=black] {$\phi_2$};
      \path[left color=black!40,right color=black!5,opacity=0.3] (-\eradiusx,\deltay) arc (180:360:{{\eradiusx}} and {{\eradiusy}}) -- (\eradiusx,-\deltay) arc(360:180:{{\eradiusx}} and {{\eradiusy}}) -- (-\eradiusx,\deltay);
    \end{scope}
  \end{tikzpicture}
    \caption{Boundary anchored geodesics making up the kinematic space for the AdS$_3$ geometry dual to a conformally transformed vacuum state on the left and the $\ZZ_3$ conical defect dual to an excited state on the right.}
    \label{fig:geodesics}
\end{figure}
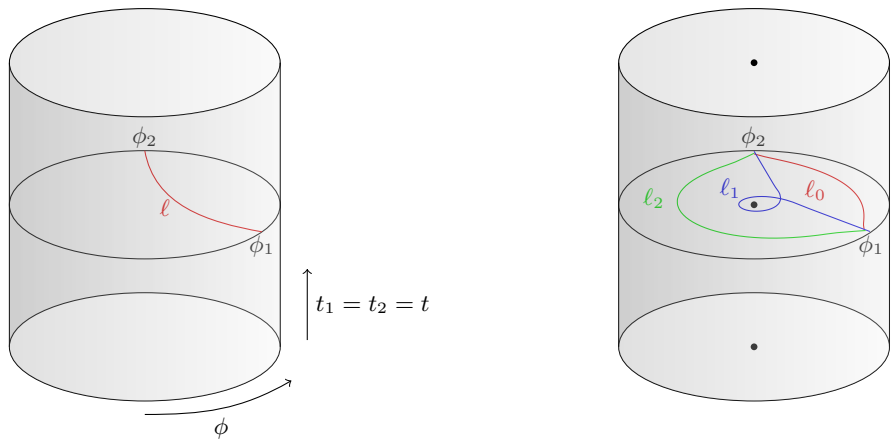

We will construct this gravity dual using techniques inspired by mathematical tools from integral geometry known under the name of kinematic space~\cite{Santalo}.
In our setup the kinematic space is the space of all geodesics anchored on the asymptotic boundary~\footnote{Note that this includes winding geodesics if the spacetime has non-trivial topology, see Fig.~\ref{fig:geodesics}.}.
This auxiliary space has been used previously to reformulate geometric objects in asymptotically AdS$_3$ spaces as functionals on the kinematic space, see e.g.~\cite{Czech,Czech:2014ppa,Czech,Czech:2015kbp,deBoer:2015kda,Czech:2016xec,deBoer:2016pqk,Czech:2017zfq,Cresswell:2017mbk,Zhang:2016evx,Abt:2017pmf,Abt:2018ywl,Czech:2019hdd}.
In particular, complexity measures outside the realm of complexity geometry have been explored in the context of the kinematic space in~\cite{Czech:2017ryf,Chen:2020nlj}.
By the Ryu-Takayanagi formula, which associates the length of the geodesics in the kinematic space with the CFT entanglement entropy, the kinematic space formulation of the problem allows a derivation of duals to bulk geometric objects in terms of boundary entanglement data.
Here, we will do the reverse: we will use the kinematic space to map the Fubini-Study cost function \eqref{eq:Fubini-Study-line-element} - a boundary quantity - to a geometric object in the bulk defined by its formulation as a kinematic space functional.

Let us first describe the kinematic space in more detail.
Each geodesic is specified by its two endpoints $(z_1,\bar{z}_1)$ and $(z_2,\bar{z}_2)$.
The length of this geodesic in our geometry dual to a quantum circuit is given by
\begin{equation}
  \ell = \log\left[\frac{\sin((f(t_2,z_2)-f(t_1,z_1))/2)\sin((\bar{f}(t_2,\bar{z}_2)-\bar{f}(t_1,\bar{z}_1))/2)}{\epsilon_\text{UV}^2}\right],
  \label{eq:geodesic-length}
\end{equation}
where $f,\bar{f}$ parametrize the conformal transformations which define the circuit and $\epsilon_\text{UV}$ is a UV cutoff.

The goal is now to find a relation between the geodesic length $\ell$ and the two-point function of the Hamiltonian density whose integral gives the Fubini-Study cost function \eqref{eq:Fubini-Study-cost-function}. For the trivial circuit $f(t,z)=z$, $\bar{f}(t,\bar{z})=\bar{z}$, it is easy to express connected two-point correlators of the energy-momentum tensor in terms of $\ell$,
\begin{equation}
  \langle T(z_1) T(z_2) \rangle = \frac{c}{32}\frac{1}{\sin((z1-z2)/2)^4} = \frac{c}{2}(\partial_{z_1}\partial_{z_2}\ell)^2.
\end{equation}
This leads to the following connected two-point function of the Hamiltonian density,
\begin{equation}
  \begin{aligned}
    \langle \cH(z_1,\bar{z}_1) \cH(z_2,\bar{z}_2) \rangle &= \langle T(z_1)T(z_2) \rangle + \langle \bar{T}(\bar{z}_1)\bar{T}(\bar{z_2}) \rangle\\
    &= \frac{c}{2}\left[(\partial_{z_1}\partial_{z_2}\ell)^2 + (\partial_{\bar{z}_1}\partial_{\bar{z}_2}\ell)^2\right]\\
    &= \frac{c}{4}\left[(\partial_{\phi_1}\partial_{t_2}\ell)(\partial_{t_1}\partial_{\phi_2}\ell) + (\partial_{\phi_1}\partial_{\phi_2}\ell)(\partial_{t_1}\partial_{t_2}\ell)\right].
  \end{aligned}
\end{equation}
For our circuit we have to allow arbitrary $f$ and $\bar{f}$.
In this case, after some algebra we find the following relation between the connected two-point function of the Hamiltonian density $\cH$ and the geodesic length $\ell$,
\begin{gather}
  \sqrt{g_{(0)}(t_1,\phi_1)}\sqrt{g_{(0)}(t_2,\phi_2)}\langle \cH(t_1,\phi_1) \cH(t_2,\phi_2) \rangle \nonumber
  \\
 = \partial_{\phi_1}f_1\partial_{t_1}f_1\partial_{\phi_2}f_2\partial_{t_2}f_2\langle T(f_1)T(f_2) \rangle + \partial_{\phi_1}\bar{f}_1\partial_{t_1}\bar{f}_1\partial_{\phi_2}\bar{f}_2\partial_{t_2}\bar{f}_2\langle \bar{T}(\bar{f}_1)\bar{T}(\bar{f}_2) \rangle = \cF_\text{bulk}
  \label{eq:Hamiltonian-correlator}   
\end{gather}
where
\begin{equation}
  \cF_\text{bulk} = \frac{c}{4}\biggl[
    (\partial_{\phi_1}\partial_{\phi_2}\ell)(\partial_{t_1}\partial_{t_2}\ell) + (\partial_{\phi_1}\partial_{t_2}\ell)(\partial_{t_1}\partial_{\phi_2}\ell)
    - \frac{1}{2}g^{(0)}_{t_1\phi_1}g^{(0)}_{t_2\phi_2} g_{(0)}^{ij}(t_1,\phi_1)g_{(0)}^{kl}(t_2,\phi_2)(\partial_i\partial_k\ell)(\partial_j\partial_l\ell)
    \biggr].
  \label{eq:definition-F_bulk}
\end{equation}
Therefore, we have found a bulk dual to the Fubini-Study cost function,
\begin{equation}
  \begin{aligned}
     F_\text{FS}(t) &= \int d\phi_1 \int d\phi_2 \sqrt{g_{(0)}(t,\phi_1)}\sqrt{g_{(0)}(t,\phi_2)} \langle \cH(t,\phi_1) \cH(t,\phi_2) \rangle\\
    = F_\text{bulk}(t) &= \int d\phi_1 \int d\phi_2\, \cF_\text{bulk},
  \end{aligned}
  \label{eq:bulk-dual-FS-distance}
\end{equation}
expressed in terms of geodesic lengths.
This mapping between a natural measure of distance between states and a purely geometric object in the bulk is our first main technical result.
As~\eqref{eq:bulk-dual-FS-distance} was derived from the AdS/CFT dictionary, it is valid for any quantum circuit built out of conformal transformations.
Note that despite the geodesic length $\ell$ being UV divergent, \eqref{eq:bulk-dual-FS-distance} is UV finite.
This is to be expected, since applications of exponents of smeared local operators (in our case, the energy-momentum tensor) are not expected to alter the ultraviolet behaviour of the states. Indeed, the Fubini-Study distance was shown in~\cite{Flory:2020dja,Flory:2020eot} to exhibit UV finiteness for the circuits used here\footnote{To be precise, it is necessary to choose a regularization procedure for the integral around the point $\phi_1=\phi_2$ where the two energy-momentum tensor insertions collide \cite{Flory:2020dja,Flory:2020eot}. This regularization is implicit in writing the Fubini-Study distance as the integral over local operators and does not affect the final result.}.
 
Interestingly, the volume $V$ of a constant time slice of pure AdS$_3$ from the ``complexity=volume'' proposal can be also obtained from the kinematic space formulation in the trivial circuit $f(t,z)=z$, $\bar{f}(t,\bar{z})=\bar{z}$.
It is given by a formula similar to \eqref{eq:bulk-dual-FS-distance} \cite{Abt:2017pmf,Abt:2018ywl},
\begin{equation}
  V = \int d\phi d\rho \sqrt{g_\text{ind}(\phi,\rho)} \propto \int d\phi_1 \int d\phi_2\, \ell\,\partial_{\phi_1}\partial_{\phi_2}\ell,
  \label{eq:volume-kinematic-space}
\end{equation}
where $g_\text{ind}$ is the induced metric on the constant time slice in the bulk and $\rho$ the AdS$_3$ radial coordinate.
Both \eqref{eq:volume-kinematic-space} and \eqref{eq:bulk-dual-FS-distance} are quadratic in the geodesic length $\ell$ but differ in the structure of derivatives applied to $\ell$.
Because $\ell$ appears without a derivative in \eqref{eq:volume-kinematic-space}, the volume is UV divergent.

\subsection*{Conical defects}
In fact, while we have derived the formula \eqref{eq:bulk-dual-FS-distance} only for circuits comprising states that are conformal transformations of the vacuum state, it applies to a much wider set of cases.
In particular, we find that \eqref{eq:bulk-dual-FS-distance} also holds for conformal transformations of primary states with conformal weight $h = \frac{c}{24}(1-1/n^2)$ where $n \in \NN$.
For excited primary states, the connected two-point function of the energy-momentum tensor is given by
\begin{equation}
   \bra{h}T(z_1)T(z_2)\ket{h} = \frac{c}{32}\frac{1}{\sin((z1-z2)/2)^4} - \frac{h}{2}\frac{1}{\sin((z1-z2)/2)^2}
\end{equation}
These states are dual to conical defects arising from a $\ZZ_n$ identification of pure AdS$_3$.
In this spacetime, there are $n$ geodesics connecting two boundary points $(t_1,\phi_1)$, $(t_2,\phi_2)$ with winding numbers $w=0,...,n-1$ (see \figref{fig:geodesics}) and length
\begin{equation}
    \ell_w = \log\left[\frac{\sin\left(\left(f(t_2,z_2)-f(t_1,z_1)+2\pi w\right)/2n\right)\sin\left(\left(\bar{f}(t_2,\bar{z}_2)-\bar{f}(t_1,\bar{z}_1)-2\pi w\right)/2n\right)}{\epsilon_\text{UV}^2}\right],
  \label{eq:geodesic-length-conical-defect}
\end{equation}
Therefore, the kinematic space consists of $n$ sectors with fixed winding number $w$ and to obtain a bulk dual to the connected two-point function of the Hamiltonian density, we have to sum over all $w$,
\begin{equation}
  \begin{aligned}
  &\sqrt{g_{(0)}(t_1,\phi_1)}\sqrt{g_{(0)}(t_2,\phi_2)}\bra{h} \cH(t_1,\phi_1) \cH(t_2,\phi_2) \ket{h}\\
  & = \partial_{\phi_1}f_1\partial_{t_1}f_1\partial_{\phi_2}f_2\partial_{t_2}f_2\bra{h} T(f_1)T(f_2) \ket{h} + (c.c)\\
  & = \begin{aligned}[t]\sum_{w=0}^{n-1}\frac{c}{4}\biggl[&
    (\partial_{\phi_1}\partial_{\phi_2}\ell_w)(\partial_{t_1}\partial_{t_2}\ell_w) + (\partial_{\phi_1}\partial_{t_2}\ell_w)(\partial_{t_1}\partial_{\phi_2}\ell_w)\\
    &- \frac{1}{2}g^{(0)}_{t_1\phi_1}g^{(0)}_{t_2\phi_2} g_{(0)}^{i_1j_1}g_{(0)}^{k_2l_2}(\partial_{i_1}\partial_{k_2} \ell_w)(\partial_{j_1}\partial_{l_2} \ell_w)\biggr].
    \end{aligned}
  \end{aligned}
  \label{eq:bulk-dual-Hamiltonian-correlator-primary-state}
\end{equation}
Inserting this into \eqref{eq:bulk-dual-FS-distance} yields again a geometric expression for the Fubini-Study cost function \eqref{eq:Fubini-Study-cost-function}.
This determines the gravity dual to the Fubini-Study distance for any two pure states related by conformal transformations (i.e.~any two states in the same Verma module).

\subsection*{BTZ black holes}
Apart from conical defects, the expression \eqref{eq:bulk-dual-FS-distance} also correctly reproduces the connected two-point function of the Hamiltonian in a thermal state dual to a BTZ black hole or the TFD state
\begin{equation}
\label{eq.TFDdef}
\ket{\mathrm{TFD}(t)} = \frac{1}{\sqrt{Z(\beta)}}\sum_n e^{-i E_{n} t}e^{-\beta E_n/2}\ket{E_n}_L\ket{E_n}_R
\end{equation}
dual to the two-sided BTZ geometry, where the sum runs over all energy eigenstates.
In this case, there are multiple geodesics to consider.
The length of a geodesic stretching between two-points on the asymptotic boundary of the one-sided BTZ black hole is given by
\begin{equation}
  \ell = \log\left[\frac{\cosh(2\pi(\phi_1-\phi_2)/\beta) - \cosh(2\pi(t_1-t_2)/\beta)}{\epsilon_\text{UV}}\right]
\end{equation}
while geodesics between two different asymptotic boundaries in the maximally extended two-sided BTZ geometry have length
\begin{equation}
  \ell = \log\left[\frac{\cosh(2\pi(\phi_1-\phi_2)/\beta) + \cosh(2\pi(t_1+t_2)/\beta)}{\epsilon_\text{UV}}\right].
\end{equation}
For geodesics with winding numbers, simply set $\phi_1-\phi_2 \to \phi_1-\phi_2 + 2\pi w$.
Applying \eqref{eq:Hamiltonian-correlator} and \eqref{eq:definition-F_bulk}, integrating over $\phi_1$ and $\phi_2$ as well as summing over all possible winding numbers\footnote{The sum over winding numbers is equivalent to integrating $\chi=(\phi_1+\phi_2)/2$ from 0 to $2\pi$ and $\psi=(\phi_1-\phi_2)/2$ from 0 to $\infty$.} leads also in this case to a result proportional to the thermal two-point function
\begin{equation}
  \langle H^2 \rangle_\beta - \langle H \rangle_\beta^2 = \frac{\partial_\beta^2 Z(\beta)}{Z(\beta)} - \left(\frac{\partial_\beta Z(\beta)}{Z(\beta)}\right)^2 = \frac{2c\pi^2}{3\beta^3},
  \label{eq:thermal-two-point function}
\end{equation}
where $Z(\beta) = \Tr[e^{-\beta H}] = \exp\left(\frac{c}{12}\frac{4\pi^2}{\beta}\right)$ is the thermal partition function.
For the two-sided black hole, the Hamiltonian $H=H_L + H_R$ is the sum of the two Hamiltonians on the left and right asymptotic boundaries and its connected two-point function in the TFD state is given by four times the result of \eqref{eq:thermal-two-point function}.
In the bulk, these four contributions come from the four possibilities of placing boundary points $(\phi_1,t_1)$, $(\phi_2,t_2)$ on the two asymptotic boundaries.

We note two small subtleties concerning this result.
First, in the case where the integral in \eqref{eq:bulk-dual-FS-distance} runs over the endpoints of geodesics on the same asymptotic boundary, the integral is formally divergent due to infinities at colliding operator insertion points where $\phi_1=\phi_2$.
A consistent result is obtained through regularization.
By restricting $\phi_1-\phi_2$ to be greater than some value $\tilde\epsilon$ we obtain a result where the two-point function of the Hamiltonian emerges at order $O(\tilde\epsilon^0)$.
Second, in the case where the integral runs over geodesic lengths on different asymptotic boundaries, the time derivatives in \eqref{eq:definition-F_bulk} have to be taken with respect to~the Killing time which is given by $t_1$, respectively~$-t_2$, on the left, respectively~right, asymptotic boundary in order to obtain the correct sign.

In summary, we have found a geometric dual, expressed in terms of geodesic lengths, to the Fubini-Study distance between two states related by infinitesimal time-evolution.
This expression applies to all asymptotically AdS$_3$ spacetimes without bulk matter fields, i.e.~Ba{\~n}ados geometries corresponding to conformal transformations of the vacuum state as well as excited primary states  and the BTZ black hole corresponding to a thermal state.
This maps a cost function of a field theory complexity measure into a purely geometric quantity in the dual field theory, opening the door to studying gravity duals to computational complexity from first principles.

\section{From cost to complexity}
\label{sec:complexity}

So far, we have discussed the map between the boundary cost and bulk geometry. With complexity arising from the optimization of the cost, as encapsulated by Eq.~\eqref{eq.complexitydef}, in the present section we want to apply our framework to study bulk complexity completely ab initio.

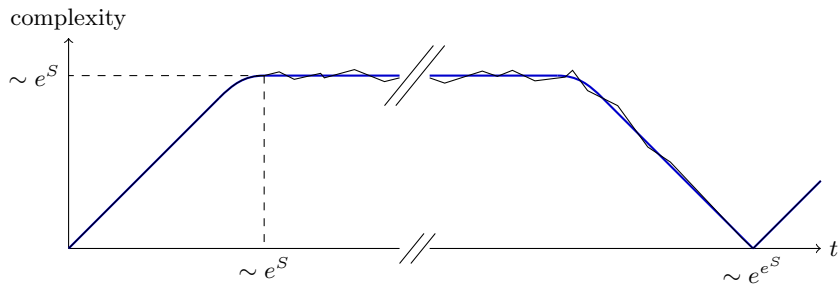
\begin{figure}
    \centering
    \begin{tikzpicture}
        \draw (0,0) -- (4.4,0);
        \draw[->] (4.8,0) -- (10,0) node[right] {$t$};
        \draw (4.4,-0.2) -- (4.725,0.2);
        \draw (4.55,-0.2) -- (4.875,0.2);
        \draw[->] (0,0) -- (0,2.8) node[above] {\small complexity};
        \draw[color=blue!80!black,thick] (0,0) -- (2,2);
        \draw[color=blue!80!black,thick] (2,2) to[out=45,in=180] (2.6,2.3);
        \draw[color=blue!80!black,thick] (2.6,2.3) -- (4.4,2.3);
        \draw (4.2,1.9) -- (4.85,2.7);
        \draw (4.35,1.9) -- (5.0,2.7);
        \draw[color=blue!80!black,thick] (4.8,2.3) -- (6.5,2.3);
        \draw[color=blue!80!black,thick] (6.5,2.3) to[out=0,in=135] (7.1,2);
        \draw[color=blue!80!black,thick] (7.1,2) -- (9.1,0);
        \draw[color=blue!80!black,thick] (9.1,0) -- (10,0.9);
        \draw[dashed] (2.6,2.3) -- (0,2.3) node[left] {$\sim e^S$};
        \draw[dashed] (2.6,2.3) -- (2.6,0) node[below] {$\sim e^S$};
        \draw (9.1,0) node[below] {$\sim e^{e^S}$};
        \draw[thin] (0,0) -- (2,2) to[out=45,in=180] (2.6,2.3) -- (2.8,2.35) -- (3,2.25) -- (3.35,2.33) -- (3.4,2.27) -- (3.8,2.38) -- (4.2,2.22) -- (4.4,2.27);
        \draw[thin] (4.8,2.27) -- (5.0,2.2) -- (5.5,2.36) -- (5.7,2.29) -- (5.9,2.37) -- (6.2,2.23) -- (6.6,2.28) -- (6.7,2.37) -- (6.9,2.1) -- (7.3,1.9) -- (7.7,1.35) -- (8.0,1.15) -- (8.5,0.62) -- (8.95,0.14) -- (9.1,0) -- (10,0.9);
    \end{tikzpicture}
    \caption{Time evolution of computational complexity as conjectured by \cite{Susskind:2014rva}.}
    \label{fig:time-evolution-complexity}
\end{figure}

The main question we are aiming to answer is the following.
It has been conjectured that the computational complexity in quantum systems which describe AdS black holes evolves universally (i.e. for any reasonable definition of complexity) in a particular way with time. It is supposed to increase linearly until an exponentially long time scale in the black hole entropy $S$, saturates over a duration of a doubly exponentially long time scale until it decreases again at the quantum recurrence time $t_\text{rec}$ and the process starts anew \cite{Susskind:2014rva,Stanford:2014jda,Brown:2017jil,Susskind:2018pmk}\footnote{See also \cite{Haferkamp:2021uxo,Oszmaniec:2022srs} for a proof for randomized quantum circuits with a finite number of qubits and \cite{Iliesiu:2021ari} for a bulk quantum generalization of the ``complexity=volume'' proposal that exhibits a plateau at very late times}. We illustrate this expectation in \figref{fig:time-evolution-complexity}. Our goal for this section is to study which of these features can be reproduced in our approach.

To answer this question, let us first compute the total cost 
\begin{equation}
  F_\text{FS,tot}(t) = \int_0^t dt' F_\text{FS}(t')
  \label{eq:Fubini-Study cost}
\end{equation}
for a particular circuit that computes the time-evolved TFD state $\ket{\mathrm{TFD}(t)}$ dual to the BTZ black hole in question, without minimizing over different circuits to obtain the complexity.
The results for $F_\text{FS,tot}$ then provide an upper bound for the complexity.
There is one particular circuit for which this is particularly easy to do this and that is the time evolution of the BTZ black hole itself where $\ket{\mathrm{TFD}(t)}$ is created simply by ordinary time evolution with $H=H_L+H_R$ and $H_{L,R}=L_0+\bar{L}_0$ from the reference state $\ket{\mathrm{TFD}(0)}$.
In that case we indeed find a linear increase with $t$ in $F_\text{FS,tot}(t)$ due to $F_\text{FS}(t)$ being constant in $t$.
Therefore, if the conjecture on the linear increase of complexity is true, then the optimal circuit for constructing the time-evolved TFD state is not far away from ordinary time evolution for less than exponential times.

Similar arguments lead to the dip at the recurrence time.
By also allowing circuits which time evolve reversely (i.e.~evolution with $-H$ instead of $H$) it is obvious that the complexity is upper bounded by \begin{equation}
    C_\text{FS}(t) \leq \min\bigl(\int_0^t dt' F_\text{FS}(t'), \int_t^{t_\text{rec}} dt' F_\text{FS}(t')\bigr)
\end{equation}
and therefore must decrease to zero again at $t=t_\text{rec}$.

To show that ordinary time evolution is indeed optimal, we now optimize the total cost over all circuits that connect $\ket{\mathrm{TFD}(0)}$ with $\ket{\mathrm{TFD}(t)}$ via applying conformal transformations,
\begin{equation}
    C_\text{FS}(t) = \min F_\text{FS,tot}(t).
\end{equation}
The minimization procedure proceeds by performing a variation of $F_\text{FS,tot}(t)$ with respect to~$f(t,z)$ and solving the resulting equations of motion to determine the path $f(t,z)$ of least total cost.
For simplicity, we restrict to conformal transformations acting only on one boundary of the wormhole.\footnote{Note that coupling the both boundaries/the both underlying quantum field theories, as, for example, in~\cite{Gao:2016bin,Chapman:2018hou}, goes significantly beyond the algebraic setup of ours.}

The Fubini-Study cost function in this situation of interest is given by
\begin{equation}
    \begin{aligned}
        F_\text{FS}(t) &= \bra{\psi(t)}Q(t)^2\ket{\psi(t)} - \bra{\psi(t)}Q(t)\ket{\psi(t)}^2\\
        &= \bra{\mathrm{TFD}(0)}\tilde Q(t)^2\ket{\mathrm{TFD}(0)} - \bra{\mathrm{TFD}(0)}\tilde Q(t)\ket{\mathrm{TFD}(0)}^2.
    \end{aligned}
    \label{eq:Fubini-Study-TFD}
\end{equation}
Here, the circuit Hamiltonian $Q(t)$ is defined in \eqref{eq:circuit-Hamiltonian} while $\tilde Q(t)$ is its conformal transformation defined by letting the path ordered exponential from \eqref{eq.circuit} act on $Q(t)$ instead of $\ket{\mathrm{TFD}(0)}$ in \eqref{eq:Fubini-Study-TFD}.
From the well-known transformation law of the stress-energy tensor, $\tilde Q(t)$ can be written as
\begin{equation}
    \tilde Q(t) = \int d\phi \left(1+\frac{\dot f(t,z)}{f'(t,z)}\right) \left[T(z) - \frac{c}{12}\{f,z\}\right]
\end{equation}
using $\epsilon(t,z) = \dot f(t,F(t,z)) + f'(t,F(t,z))$ from \eqref{eq:epsilon-coefficient}.
The notation is such that $F(t,z)$ is the inverse of $f(t,z)$, i.e.~$f(t,F(t,z)) = z$, while $\dot f$ is the derivative w.r.t.~the first argument of $f$ and $f'$ the one w.r.t.~the second argument.
The thermal two-point function of the energy-momentum tensor can be obtained by the conformal Ward identity \cite{Eguchi:1986sb},
\begin{equation}
    \begin{aligned}
        \bra{\mathrm{TFD}(0)}T(z_1)T(z_2)\ket{\mathrm{TFD}(0)} = &\frac{c}{24}\wp''\left(\frac{z_1-z_2}{2\pi}\right) + 2\frac{2\pi i \partial_\tau Z(\tau)}{Z(\tau)}\left(\wp\left(\frac{z_1-z_2}{2\pi}\right)+2\eta_1\right)\\
        &+ \frac{(2\pi i\partial_\tau)^2Z(\tau)}{Z(\tau)}-\left(\frac{2\pi i\partial_\tau Z(\tau)}{Z(\tau)}\right)^2,
    \end{aligned}
\end{equation}
where $Z(\tau)$ is the (holomorphic part of) the partition function and $\wp(z)$ denotes the Weierstraß elliptic function with associated parameter $\eta_1$.
From periodicity in $\phi$, we find a Fubini-Study cost of the form
\begin{equation}
    F_\text{FS} = \sum_{n \geq 0} \alpha_n \gamma_n \gamma_{-n},
    \label{eq:Fubini-Study-Fourier-expansion}
\end{equation}
where $\alpha_n$ are positive numbers and $\gamma_n$ are the Fourier coefficients of $\gamma = 1 + \dot f/f'$.
Inserting \eqref{eq:Fubini-Study-Fourier-expansion} into \eqref{eq:Fubini-Study cost} and varying w.r.t.~$f_n$, the Fourier coefficients of $f$, determines the optimal path with minimum total cost.
To solve the resulting equations of motion we expand in a perturbation parameter $\sigma$, $f(t,z) = z + \sigma f^{(1)}(t,z) + \sigma^2 f^{(2)}(t,z) + O(\sigma^3)$.
Written in a polar decomposition for the Fourier coefficients $f^{(k)}_n(t) = |f^{(k)}_n(t)| e^{i\theta^{(k)}_n(t)}$, this gives to leading order in $\sigma$ the equations of motion
\begin{equation}
    \begin{aligned}
        &\ddot f^{(1)}_0 = 0 \quad \text{for $n=0$}\\
        &\alpha_0 n \partial_t |f^{(1)}_{n}|^2 = 0, \quad \alpha_n |\ddot f^{(1)}_n| + 2\alpha_0 n\dot\theta^{(1)}_n|f^{(1)}_n| = 0 \quad \text{for $n>0$.}
    \end{aligned}
    \label{eq:eom-leading-order}
\end{equation}
For the target state $\ket{\mathrm{TFD}(t_f)}$, the boundary conditions for these equations are $f(t=0) = f(t=t_f) = z$.
Therefore, we find from \eqref{eq:eom-leading-order} that $f^{(1)}_n = 0$ for all $n$.
As the first order contribution in $\sigma$ to $f(t,z)$ vanishes, we deduce that the second order contribution vanishes as well (simply replace $\sigma^2 \to \sigma$ to get the same equation of motion for $f^{(2)}_n$ as for $f^{(1)}_n$).
Hence, order by order in perturbation theory we find that time evolution by non-trivial conformal transformations is more expensive in our setup than ordinary time evolution with $H = L_0 + \bar{L}_0$.
Thus, the computational complexity increases or decreases linearly in the regime where perturbation theory is applicable, i.e.~at times close to zero or close to a multiple of the recurrence time.

In the complexity geometry picture of \cite{Nielsen}, contributions which might be invisible in perturbation theory can in particular appear at conjugate points in the complexity geometry. In previous studies of computational complexity in chaotic systems, the appearance of conjugate points has lead to a saturation of computational complexity, see e.g.~\cite{Balasubramanian:2019wgd,Balasubramanian:2021mxo}.
In our setup, the notion of conjugate points on a geodesic in a Finsler geometry as used by \cite{Nielsen} is not applicable  because the squared Fubini-Study distance is positively homogeneous of degree two under $H \to \alpha H$ instead of degree one as required for a Finsler metric.
Nevertheless, similar effects which our perturbative calculation cannot capture might still lead to a saturation in our case.

\section{The Fubini-Study cost function and the switchback effect}
\label{sec:switchback}

Given that the bulk dual to the Fubini-Study cost derived in \secref{sec:bulk-dual-Fubini-Study} in the BTZ geometry correctly reproduces the thermal two-point function, we now study what happens if bulk matter fields enter the picture.
In particular, we investigate BTZ geometries perturbed by adding shockwaves.
In this situation computational complexity shows a striking time dependence where an exponential growth regime turns into a linear one at the scrambling time, a phenomenon known as the switchback effect \cite{Stanford:2014jda,Susskind:2014jwa}.

We find that the geometric expression $F_\text{bulk}$ from eq.~\eqref{eq:definition-F_bulk}, which we found to be dual to the Fubini-Study distance, is sensitive to the switchback effect but that the Fubini-Study cost is not.
This shows that the equality between the Fubini-Study distance and $F_\text{bulk}$ is not applicable to geometries sourced by bulk matter fields (and indeed there is no reason to expect it to).
Nevertheless, the fact that $F_\text{bulk}$ can probe the switchback effect implies that this quantity shows the features generally expected from a cost function for a holographic complexity measure~\cite{Belin:2021bga,Belin:2022xmt}: linear growth at late times in the BTZ wormhole geometry and an imprint of the switchback effect \cite{Susskind:2014rva,Susskind:2014jwa,Stanford:2014jda}.

Let us now briefly review the switchback effect on the field theory side.
In the computational complexity setup, the perturbation generating the shockwave in the dual bulk picture is implemented by applying a precursor operator which \emph{in the Schrödinger picture} is given by $U^\dagger(t_W) W U(t_W)$.
The standard, discrete gate-counting complexity of this operator on its own can be estimated by a simple infection model \cite{Stanford:2014jda,Susskind:2018pmk}.
In this model, the operator $W$ acts on a single qubit and is counted as a single gate.
The complexity of the precursor $U^\dagger(t_W) W U(t_W)$ is then given by one plus the number of gates in $U^\dagger(t_W)$ and~$U(t_W)$ minus the number of gates that cancel between $U^\dagger(t_W)$ and~$U(t_W)$.
How many such cancellations occur is determined by an infection model: by acting with $W$ on a qubit, this qubit becomes infected and the infection spreads throughout the quantum system when an infected qubit couples to a non-infected one by two- or more qubit gates in $U(t_W)$.
Gates inside $U(t_W)$ and $U^\dagger(t_W)$ cancel if they act on non-infected qubits.
This leads to a complexity of the precursor operator given by \cite{Stanford:2014jda,Susskind:2018pmk}
\begin{equation}
  C \propto K\log\left(1+e^{\frac{2\pi}{\beta}(t_W-t_*)}\right),
  \label{eq:complexity-switchback-effect}
\end{equation}
where $K$ is the number of qubits and $t_*$ is the scrambling time, given by $t_* = \frac{\beta}{2\pi}\log c$ in two-dimensional CFTs \cite{Roberts:2014ifa}. There is a characteristic time delay in the complexity growth:  for $W \neq \mathbf{1}$ there is first an exponential increase in $t_W$ for small $t_W$ followed by a linear growth regime \footnote{It is important to note that this behaviour is observed for the $t_W$ dependence, the parameter of the precursor.
The model does not capture subsequent time evolution $U(t)\ket{\psi(0)} = U(t) U^\dagger(t_W)WU(t_W)\ket{\mathrm{TFD}(0)}$ occuring after the precursor has been applied onto the TFD state and hence the complexity is only the complexity of the precursor operator itself.}.

In the context of two-dimensional conformal field theories, the operator $W$ is taken to be a heavy primary operator whose insertion at $t=t_W \to -\infty$ leads to a null shockwave propagating along the horizon \cite{Roberts:2014ifa}.
On the gravity side, the computational complexity of the precursor is then estimated by applying one of the holographic complexity proposals\footnote{At $t=0$ in order to capture the complexity of the precursor and not any subsequent time evolution.} which shows the same time-evolution behaviour \eqref{eq:complexity-switchback-effect} as obtained from the infection models described above.
This has been derived for the ``complexity=volume'' proposal in \cite{Stanford:2014jda} and for the ``complexity=action'' proposal in \cite{Brown:2015bva,Brown:2015lvg}.
This property also holds by construction for the infinite families of holographic complexity measures proposed in \cite{Belin:2021bga,Belin:2022xmt}, as it is one of their two defining features.

However, for the field theory quantity that we are considering -- the Fubini-Study cost~\eqref{eq:Fubini-Study-cost-function} and its time integral~\eqref{eq:Fubini-Study cost} -- it is clear that \eqref{eq:complexity-switchback-effect} cannot be reproduced.
It is easy to see\footnote{Let the TFD state be given by $\ket{\mathrm{TFD}} = \frac{1}{\sqrt{Z(\beta)}}\sum_n e^{-\beta E_n/2}\ket{E_n}_L\ket{E_n}_R$ and expand the operator $W$ in energy eigenstates, $W = \sum_{n,m} W_{nm}\ket{E_n}_L\bra{E_m}_L \times \mathbf{1}_R$.
Then, by an elementary calculation we see that the expectation values $\bra{\psi}H\ket{\psi}$ and $\bra{\psi}H^2\ket{\psi}$ are independent of~$t_W$ and, via~\eqref{eq:Fubini-Study-cost-function}, it settles the point.} that the Fubini-Study distance between the state $\ket{\psi} = U^\dagger(t_W) W U(t_W)\ket{\mathrm{TFD}(0)}$ and the state $e^{i H dt}\ket{\psi}$ is independent of $t_W$. In fact, this argument showing that expectation values $\bra\psi\cO\ket\psi$ are independent of $t_W$ holds for any operator $\cO$ that is diagonal in the energy eigenbasis.

Nevertheless, the geometric expression $F_\text{bulk}$ obtained in \secref{sec:bulk-dual-Fubini-Study} -- which for bulk geometries without matter fields is equal to the Fubini-Study cost function -- reproduces the switchback effect as we will show below.
This makes clear that this expression is not dual to the Fubini-Study cost function for the perturbed TFD states.

We now come to the gravitational calculation of $F_\text{bulk}$ in the shockwave geometries.
The shockwave geometries we consider are given by portions of BTZ geometries glued together along null surfaces \cite{Shenker:2013pqa}.
For a shockwave moving from the bottom left to the top right in the Penrose diagram, in Kruskal coordinates a point $(u=0,v,\phi)$ on one side is identified with $(u=0,v+h(\phi),\phi)$ on the other side where the displacement $h(\phi)$ depends on the exact form of the bulk matter concentration that sources the shockwave.
We will consider bulk matter that is concentrated along the horizon at $u=0$ in Kruskal coordinates as well as localized at $\phi=\hat\phi$ in the angular direction similar to the setup in~\cite{Roberts:2014isa}.
The periodicity $\phi \sim \phi + 2\pi$ requires the bulk energy-momentum tensor to be given by
\begin{equation}
  T_{uu} = \frac{\alpha}{2\pi G_N} \sum_{n \in \ZZ}\delta(u)\delta(\phi - \hat \phi + 2\pi n).
\end{equation}
while the other components of $T_{\mu\nu}$ vanish.
The parameter
\begin{equation}
  \alpha=2\exp\left(-\frac{2\pi}{\beta}(t_W-t_*)\right)
  \label{eq:alpha-parameter}
\end{equation}
determines the strength of the shock \cite{Stanford:2014jda}.
Solving Einstein's equations gives a metric of the form
\begin{equation}
  ds^2 = - \frac{4}{(1+uv)^2}dudv + r_H^2\frac{(1-uv)^2}{(1+uv)^2} d\phi^2 + 4 \delta(u)h(\phi)du^2,
  \label{eq:shockwave-metric}
\end{equation}
where the displacement along the horizon at $u=0$ is given by
\begin{equation}
  v \to v + h(\phi) \quad \text{with} \quad h(\phi) = \alpha \frac{\cosh(r_H(|\phi - \hat \phi| - \pi (2n-1)))}{\sinh(r_H\pi)}, \quad \pi(n-1) \leq |\phi - \hat\phi| \leq \pi n.
\end{equation}
Here, $r_H = \frac{2\pi}{\beta}$ is the horizon radius.
For a shockwave moving in the opposite direction (bottom right to top left), the gluing condition is $(u-h(\phi),v=0,\phi) \sim (u,v=0,\phi)$ and $\alpha=2\exp\left(-\frac{2\pi}{\beta}(t_W+t_*)\right)$.

\begin{figure}
  \centering
  \includegraphics[width=.3\textwidth]{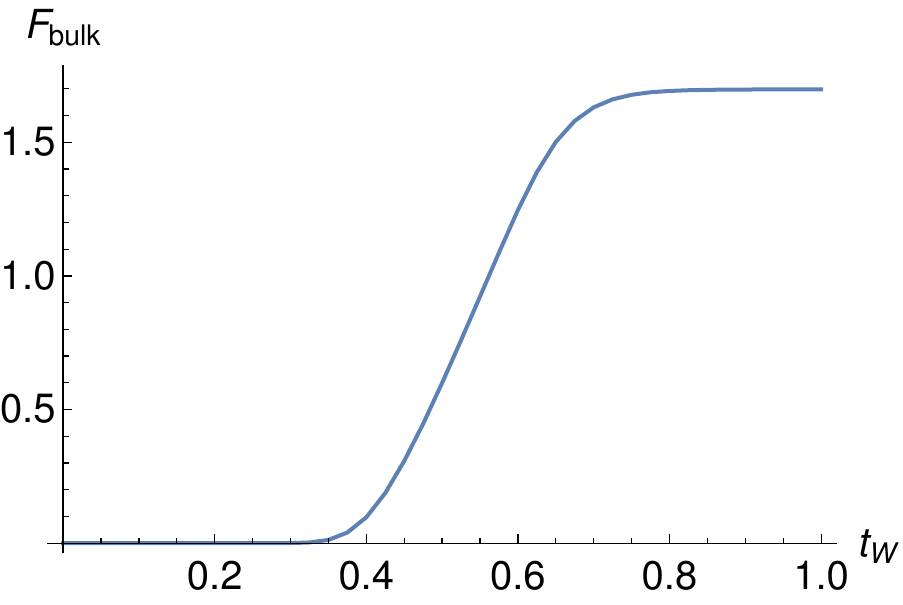}
  \includegraphics[width=.3\textwidth]{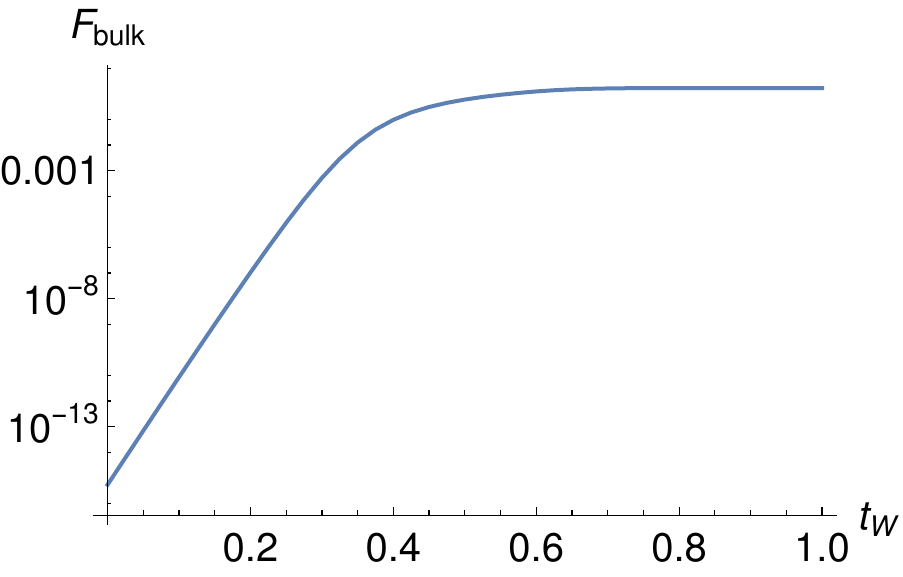}
  \includegraphics[width=.3\textwidth]{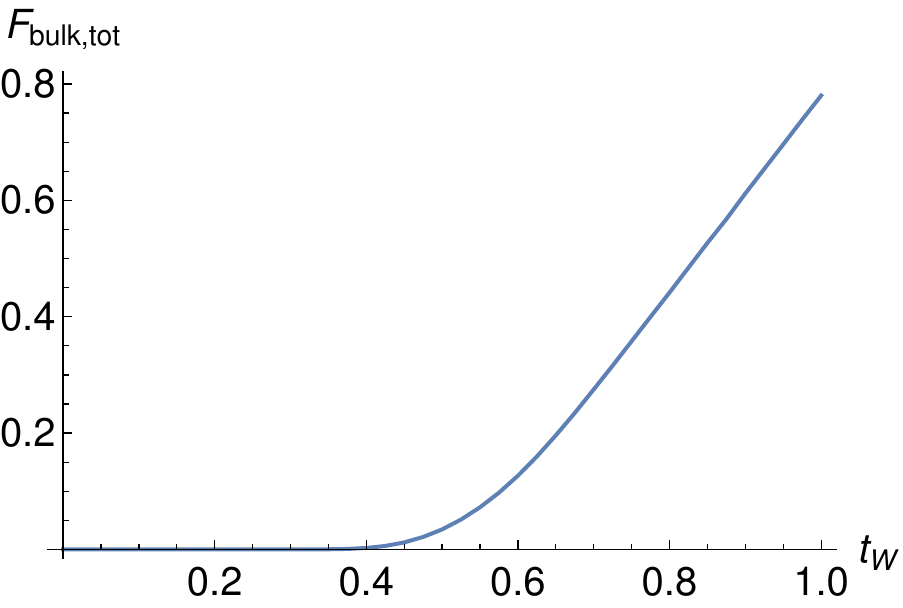}
  \caption{Left and center: the bulk cost function $F_\text{bulk}$ plotted over $t_W$ in the localized shockwave geometry with compact horizon on a linear resp.~logarithmic scale. Right: the corresponding total cost $F_\text{bulk,tot}$ (the $t_W$ integral of $F_\text{bulk}$). The plots match perfectly with the expectation from the infection models that the exponential increase, which is clearly visible in the log-scale plot in the center, turns into a linear increase at the scrambling time. All plots are evaluated for $t_*=0.5$ and $\beta = 1/4$.}
  \label{fig:Fubini-Study-localized-shockwave}
\end{figure}

In order to check if the quantity $F_\text{bulk}$ from \eqref{eq:bulk-dual-FS-distance} can reproduce \eqref{eq:complexity-switchback-effect}, we first have to derive the geodesic lengths between opposite asymptotic boundaries of the shockwave geometries.
This is easy to do with a recipe from \cite{Shenker:2013pqa}.
All asymptotically AdS$_3$ geometries can be embedded into a four-dimensional embedding space $\mathbf{R}^{2,2}$.
The geodesic length $\ell$ between two points $(T_1,T_2,X_1,X_2)$ and $(T_1',T_2',X_1',X_2')$ in the embedding space is given by
\begin{equation}
  \cosh(\ell) = T_1T_1'+T_2T_2'-X_1X_1'-X_2X_2'.
\end{equation}
The embedding space coordinates are related to the Kruskal
\begin{equation}
  ds^2 = \frac{-4dudv+r_H^2(1-uv)^2d\phi^2}{(1+uv)^2}
\end{equation}
and BTZ coordinates
\begin{equation}
  ds^2 = -(r^2-r_H^2)dt^2 + \frac{dr^2}{r^2-r_H^2} + r^2d\phi^2
  \label{eq:BTZ-metric}
\end{equation}
by
\begin{equation}
  \begin{aligned}
    T_1 &= \frac{v+u}{1+uv} = \frac{\sqrt{r^2-r_H^2}}{r_H}\sinh(r_Ht), \quad
    &T_2 = \frac{1-uv}{1+uv}\cosh(r_H\phi) = \frac{r}{r_H}\cosh(r_H\phi),\\
    X_1 &= \frac{v-u}{1+uv} = \frac{\sqrt{r^2-r_H^2}}{r_H}\cosh(r_Ht), \quad
    &X_2 = \frac{1-uv}{1+uv}\sinh(r_H\phi) = \frac{r}{r_H}\sinh(r_H\phi).
  \end{aligned}
\end{equation}
The length of a geodesic in the shockwave geometry is then obtained by adding together the length $\ell_1$ of a geodesic running from the right boundary to a point $(u=0,v=v_s,\phi=\phi_s)$ on the horizon with the length $\ell_2$ of a geodesic running from $(u=0,v=v_s+h(\phi_s),\phi=\phi_s)$ to the left boundary.
The total geodesic length is given by $\ell=\ell_1+\ell_2$ subject to the minimality conditions
\begin{equation}
    \partial_{v_s}\ell = \partial_{\phi_s}\ell = 0.
\end{equation}
Solving these equations yields
\begin{equation}
  \ell = \log\biggl[\frac{1}{2}\bigl(
      \begin{aligned}[t]
        &\cosh(2r_H\bar t) + \cosh(2r_H\Delta\phi) + \alpha e^{r_H t_1}\cosh(r_H(\phi_2 - \hat\phi-(2n-1)\pi))\\
        &+ \alpha e^{-r_H t_2}\cosh(r_H(\phi_1 - \hat\phi-(2n-1)\pi)) + \frac{1}{2} \alpha^2 e^{2r_H\Delta t} \bigr)\biggr], \quad \Delta\phi_{n-1} \lessgtr \Delta\phi \lessgtr \Delta\phi_n,
      \end{aligned}
\end{equation}
where $\phi_{1,2} = \bar{\phi}\pm\Delta\phi$, $t_{1,2}=\bar{t}\pm\Delta t$ and $\Delta\phi_n = \frac{1}{2}\log\left[\frac{\sinh(r_H(\bar t + (\bar\phi - \hat\phi - 2\pi n)))}{\sinh(r_H(\bar t - (\bar\phi - \hat\phi - 2\pi n)))}\right]$ for $1 \leq n \leq \lfloor\frac{|\bar t| + \bar\phi - \hat\phi}{2\pi}\rfloor$.
For the $\lessgtr$ comparison, the upper comparator is chosen for $\bar t > 0$ and the lower one for $\bar t < 0$.

We then insert this geodesic length into \eqref{eq:bulk-dual-FS-distance} and evaluate the resulting expressions numerically.
The results for geodesics anchored at $t=0$ show that there is a time delay in $t_W$ before $F_\text{bulk}$ saturates to the value of the Fubini-Study distance in the TFD state (see \figref{fig:Fubini-Study-localized-shockwave}).
Consequently, the total cost \eqref{eq:Fubini-Study cost} increases linearly with a time delay.
This matches perfectly with the expectations from the simple infection models for computational complexity described above: the total cost first starts from zero, then increases exponentially until the scrambling time and linearly afterwards \footnote{Note that here we are comparing a total cost on the gravity side with expectations about complexity (the minimal total cost). Therefore the match is not obvious not only because the cost function we use on the gravity side might not be sensitive to the switchback effect but also because this effect might only appear after optimizing. However, we have already determined in \secref{sec:complexity} that complexity and total cost are the same in the BTZ black hole (ordinary time evolution is the optimal path) and therefore it is reasonable to expect that the same happens in the shockwave geometries.}.
Up to numerical errors and an overall prefactor of $t_W$ and $t_*$ which is not fixed in the infection models (one may always rescale the time which it takes for one layer of the circuit to act in the infection model by a constant), the total cost is given by the right hand side of \eqref{eq:complexity-switchback-effect} when the number of qubits $K$ is identified with the central charge $c$, as in \cite{Susskind:2018pmk}.

In App.~\ref{sec:further-shockwaves}, we study further shockwave geometries with non-compact horizons and delocalized matter concentrations where we again qualitatively reproduce the switchback effect from the geometric expression $F_\text{bulk}$ although the matching to the expectations from the infection model is not as good as above.

\section{Discussion}
\label{sec:discussion}

Providing a more microscopic picture of optimal state preparation in holography, based on quantum field theory, is an important open research question that has not been yet satisfactory answered for any of the existing holographic complexity proposals. Arguably, our best bet for a definition of complexity in quantum field theory stems from~\cite{Nielsen} and is based on minimizing cost functionals for continuous quantum circuits~\eqref{eq.circuit}. Therefore, it is natural to expect that progress on understanding of the holographic complexity should be achievable along the lines of~\cite{Nielsen} upon the use of the holographic dictionary.

In the present paper we have recognized that the Fubini-Study cost~\eqref{eq:Fubini-Study-cost-function} introduced in this context in~\cite{Chapman:2017rqy} necessarily acquires a dual gravitational interpretation if one views boundary time evolution in the presence of sources as a quantum circuit. However, typically the gravitational counterpart is a complicated quantity as it depends on two point functions of local operators in non-equilibrium states. This led us to the setting of two-dimensional holographic CFTs in which ultimately due to the underlying Virasoro algebra we were able to construct the gravity dual to the Fubini-Study cost explicitly in terms of geodesic lengths between spacelike separated boundary points, see \figref{fig:geodesics}. The bulk representation of the Fubini-Study cost~\eqref{eq:bulk-dual-FS-distance} is valid for empty AdS, conical defects and the BTZ black hole geometries. We used the latter realization to show that the Fubini-Study complexity shows the expected linear growth for the time-evolved TFD state. Finally, we demonstrated that in the presence of bulk matter fields such as shockwaves the bulk quantity we introduce with~\eqref{eq:bulk-dual-FS-distance} is no longer dual to the Fubini-Study cost. This is expected, since the geometry contains backreacting matter and our novel gravitational quantity represents the Fubini-Study cost only in the universal sector of AdS$_3$ holography given by pure gravity with negative cosmological constant. Nevertheless, \eqref{eq:bulk-dual-FS-distance} considered more generally is interesting in these geometries since -- in contrast to the boundary Fubini-study cost -- the bulk quantity  \eqref{eq:bulk-dual-FS-distance} shows the switchback effect.

There are important aspects in which the geometric object on the gravity side we constructed differs from all previous holographic complexity proposals. For clarity of presentation, let us discuss them one by one.

\begin{figure}
    \centering
    \begin{tikzpicture}
        \begin{scope}
          \fill[color=blue!80!black,opacity=0.5] (2,1.2) -- (-2,1.2) to[out=-25,in=135] (0,0) to[out=45,in=205] (2,1.2);
          \draw (2,2) -- (2,-2);
          \draw (-2,2) -- (-2,-2);
          \draw[decorate,decoration={snake,amplitude=0.8}] (-2,-2) -- (2,-2);
          \draw[decorate,decoration={snake,amplitude=0.8}] (-2,2) -- (2,2);
          \draw (2,-2) -- (2,2);
          \draw (-2,-2) -- (-2,2);
          \draw (-2,-2) -- (2,2);
          \draw (-2,2) -- (2,-2);
          \fill (2,1.2) circle[radius=0.5mm] node[right] {$t=t_0$};
          \fill (-2,1.2) circle[radius=0.5mm] node[left] {$t=t_0$};
        \end{scope}
        \begin{scope}[shift={(7,0)}]
          \fill[color=blue!80!black,opacity=0.5] (2,1.2) -- (-2,1.2) -- (-2,0) -- (2,0) -- (2,1.2);
          \draw (2,2) -- (2,-2);
          \draw (-2,2) -- (-2,-2);
          \draw[decorate,decoration={snake,amplitude=0.8}] (-2,-2) -- (2,-2);
          \draw[decorate,decoration={snake,amplitude=0.8}] (-2,2) -- (2,2);
          \draw (2,-2) -- (2,2);
          \draw (-2,-2) -- (-2,2);
          \draw (-2,-2) -- (2,2);
          \draw (-2,2) -- (2,-2);
          \fill (2,1.2) circle[radius=0.5mm] node[right] {$t=t_0$};
          \fill (-2,1.2) circle[radius=0.5mm] node[left] {$t=t_0$};
          \fill (2,0) circle[radius=0.5mm] node[right] {$t=0$};
          \fill (-2,0) circle[radius=0.5mm] node[left] {$t=0$};
        \end{scope}
    \end{tikzpicture}
    \caption{Left: bulk region swept out by geodesics contributing to $F_\text{bulk}$ (defined in \eqref{eq:definition-F_bulk}). Right: union of all the bulk regions on the left for $0 \leq t \leq t_0$ contributing to the computational complexity $C_\text{FS}(t_0)$.}
    \label{fig:codimension-zero-object-swept-out-by-geodesics}
\end{figure}
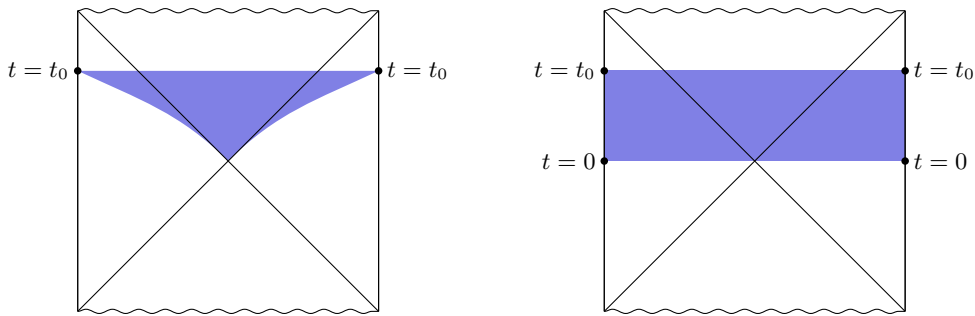

As described in \secref{sec:complexity}, we find a linearly increasing complexity (or decreasing close to the recurrence time). However, the way the linear increase arises from the bulk perspective is different in our case than in previously proposed holographic complexity measures. In the latter case, the holographic dual to computational complexity lies within a bulk region that is spacelike to the boundary time slice in which the target state is defined. The linear increase in this case is essentially due to the increasing size of the wormhole interior.
In our work, we are integrating the quantity $F_\text{bulk}$ from \eqref{eq:definition-F_bulk}, which in this case is constant, over time.
Therefore its bulk dual contains geodesics which probe the bulk region lying between the two time slices where the reference and target states are defined, see \figref{fig:codimension-zero-object-swept-out-by-geodesics}.
Another difference is that we find a computational complexity which is UV finite \footnote{Another UV finite complexity proposal generalizing CA has been studied in \cite{Mounim:2021bba}.}.
This is related to the fact that our reference states are energy eigenstates or a TFD state at $t=0$ instead of a spatially unentangled state proposed as a reference state for the earlier complexity proposals in \cite{Brown:2015lvg}. Indeed, the analyses in~\cite{Chapman:2017rqy,Jefferson:2017sdb} showed explicitly that starting with spatially disentangled state one can mimic the leading divergence of holographic complexity proposals in the setting of free quantum fields. Therefore, our bulk dual to computational complexity realises the features expected from complexity calculations in finite qubit systems in a somewhat different way than the conjectured holographic complexity proposals.
In contrast to them, however, we have derived the relation between the boundary complexity and its bulk representation from first principles.

Furthermore, our gravity expression -- while somewhat similar in spirit -- does not fit in the set of holographic complexity proposals put forward in \cite{Belin:2021bga,Belin:2022xmt} under the slogan of ``complexity=anything''. In the proposals of \cite{Belin:2021bga,Belin:2022xmt}, holographic complexity measures are constructed by integrating functionals of the bulk metric on codimension-one or codimension-zero hypersurfaces in the bulk.
These hypersurfaces are obtained by extremising functionals of the induced metric and extrinsic curvature on either the hypersurface itself in case it is codimension-one or else on the hypersurface boundaries.
Moreover, the hypersurfaces are restricted to asymptote to $t=t_0$ on the left and right asymptotic boundary of a wormhole geometry in the bulk.
As mentioned above, this restriction does not hold true for the holographic complexity measure proposed in our present work, which is defined through geodesics ending at $0 \leq t \leq t_0$ on the AdS boundary.
The geometric quantity $F_\text{bulk}$ dual to the Fubini-Study cost function, on the other hand, is naturally associated to a codimension-zero bulk region swept out by all geodesics anchored at $t=t_0$ (see \figref{fig:codimension-zero-object-swept-out-by-geodesics}) and thus constructed similarly to the complexity proposals of \cite{Belin:2022xmt}.\footnote{
To see this explicitely, note that in coordinates where the BTZ metric is given by \eqref{eq:BTZ-metric} and the corresponding coordinates $(p,q)$ for the Penrose diagram defined by $U=\text{sign}(r-r_H)\sqrt{\left|\frac{r-r_H}{r+r_H}\right|}e^{-t r_H/L^2}=\tan\left(\frac{p-q}{2}\right)$, $V=\sqrt{\left|\frac{r-r_H}{r+r_H}\right|}e^{t r_H/L^2}=\tan\left(\frac{p+q}{2}\right)$, the codimension-one surfaces bounding the codimension-zero hypersurface shown on the left hand side of \figref{fig:codimension-zero-object-swept-out-by-geodesics} are given by $t=t_0=\text{const.}$ and $q=\arctan(\sinh t_0)=\text{const.}$
As in \cite{Belin:2022xmt}, the shape of these bounding surfaces can be obtained by extremising a functional $G[\gamma,K]$ of the induced metric $\gamma$ and extrinsic curvature $K$, $\delta_X \int \sqrt{\gamma}\, G[\gamma,K]=0$ where $X$ parametrizes the location of the hypersurface.
For instance, for the upper bounding surface $q=q_0$ in \figref{fig:codimension-zero-object-swept-out-by-geodesics}, this functional can be chosen as $G[\gamma,K] = 4+R(\gamma)^2$ where $R(\gamma)$ is the Ricci scalar associated to $\gamma$.
To obtain $F_\text{bulk}$ one would then need to integrate another functional of the bulk metric inside the codimension-zero hypersurface in between the two bounding surfaces.

But even though this construction works the same way as in \cite{Belin:2022xmt}, $F_\text{bulk}$ is not a holographic complexity measure in the sense of \cite{Belin:2022xmt} due to it being constant in time instead of growing linearly at late times.}

In hindsight~\cite{Camargo:2019isp}, another approach pursuing sources in the context of state or operator preparation in quantum field theory has been the path integral optimization~\cite{Caputa:2017urj,Caputa:2017yrh,Czech:2017ryf,Bhattacharyya:2018wym,Takayanagi:2018pml}. These works, motivated by the multiscale renormalization ansatz (MERA)~\cite{Vidal:2007hda,TNRyieldsMERA}, its continuous generalization (cMERA)~\cite{Haegeman:2011uy,Nozaki:2012zj} and the quest for understanding its origins in geometric terms, pursued redundancies in Euclidean path integrals representing the same state preparation as an origin of an optimization procedure. There are two main differences between these results and our studies. The first one has to do with the fact that in our work we have a full control and understanding over our cost function, whereas in the one adopted in the path-integral optimization approach for Lorentzian circuits become problematic~\cite{Camargo:2019isp}. The second one has to do the prominent role the UV cut-off plays in the optimization procedure in the case of the path-integral optimization~\cite{Camargo:2019isp,Caputa:2020fbc} with our work using a UV-finite cost function. In particular, the latter has led to a new perspective on the path-integral optimization~\cite{Boruch:2020wax,Chandra:2021kdv,Boruch:2021hqs,Chandra:2022pgl}, which crucially involves coarse-graining CFT states~\cite{Zamolodchikov:2004ce,McGough:2016lol}. In contrast, our work is entirely phrased in the language of a local operators in CFT.

The complexity measure studied in this paper does not show the saturation behaviour expected in finite qubit systems.
From the bulk perspective, this is not particularly surprising since the bulk dual $F_\text{bulk}$ to the Fubini-Study distance we constructed is a purely classical quantity and does not take into account quantum gravity effects which were argued to lead to the saturation \cite{Susskind:2018pmk} as recently shown in two-dimensional models of quantum gravity in~\cite{Iliesiu:2021ari,Alishahiha:2022kzc,Alishahiha:2022exn}.
As it turns out, by a small modification of $F_\text{bulk}$ it is also possible to obtain a total cost which saturates at an intermediate time scale.
Summing only over geodesics in the BTZ geometry with winding numbers $w \leq w_\text{max}$ leads to an expression
\begin{equation}
  \tilde F_\text{bulk}(t) = \frac{2c\pi^2}{3\beta^3}\,\frac{\sinh\left(\frac{4\pi^2w_\text{max}}{\beta}\right)\,\left(2+\cosh\left(\frac{4\pi^2w_\text{max}}{\beta}\right)^2+3\cosh\left(\frac{4\pi^2w_\text{max}}{\beta}\right)\cosh\left(\frac{4\pi t}{\beta}\right)\right)}{\left(\cosh\left(\frac{4\pi^2w_\text{max}}{\beta}\right)+\cosh\left(\frac{4\pi t}{\beta}\right)\right)^3}
\end{equation}
which is to very good approximation constant in $t$ for $t < \pi w_\text{max}$ followed by a falloff exponential in $t$ for $t > \pi w_\text{max}$.
Integrating this in $t$ then leads to a time-dependence as expected from \figref{fig:time-evolution-complexity}, i.e.~a linear increase followed by a plateau. The CFT interpretation of this procedure is unclear\footnote{The modified cost function $\tilde F_\text{bulk}$ does not correspond to the Fubini-Study distance which for any quantum system in a thermal state evolving with a constant Hamiltonian is constant as well and thus always leads to a linearly increasing total cost.}, as but we bring it up nevertheless as a possibly interesting model of late time holographic complexity dynamics.

Finally, we want to emphasize that the results of \secref{sec:bulk-dual-Fubini-Study} relate the Fubini-Study distance to the entanglement entropy in two-dimensional holographic conformal field theories.
As the expression \eqref{eq:bulk-dual-FS-distance} for the Fubini-Study distance involves the lengths of geodesics, which are dual to entanglement entropies~\cite{Ryu:2006bv,Hubeny:2007xt,Lewkowycz:2013nqa,Dong:2016hjy}, the Fubini-Study distance is determined entirely in terms of the entanglement structure of the boundary state.
Note that for states dual to conical defects and black holes, this relation naturally involves geodesics with non-zero winding number which are dual to generalized notions of entanglement entropy that account for entanglement between different fields as well as between spatial degrees of freedom, also known as entwinement~\cite{Balasubramanian:2014sra,Balasubramanian:2016xho,Balasubramanian:2018ajb,Gerbershagen:2021gvc}.
Interestingly, these relations provide some tentative support for the idea that the modified cost function $\tilde F_\text{bulk}(t)$ defined by restricting the winding number to be smaller than some maximum can probe bulk quantum corrections as such restrictions on the maximum winding number were observed in \cite{Gerbershagen:2021gvc} and attributed to finite central charge effects which disappear in the limit were the gravity theory becomes classical.

\section{Outlook}
\label{sec:outlook}
The construction of the bulk dual to quantum circuits allows us to derive holographic duals to CFT cost functionals and vice versa from first principles. A natural next step is to expand the dictionary between CFT cost functionals and bulk quantities beyond the instance we identified and solved in the present work. One natural direction along these lines can build on~\cite{Chagnet:2021uvi} to generalize our approach to the global part of the conformal group in an arbitrary number of dimensions. Another important direction is to include additional sources on the CFT size beyond the metric, which couples to the energy-momentum tensor. Including primary operators into the picture is particularly important, as it will allow for a better understanding of the switchback effect using an entirely controllable holographic cost setup. Also, this will allow to leave the kinematic confines of the conformal group and study cost and complexity relevant to strongly-coupled theories, rather than shared by all CFTs.

By construction, the gravity dual to the circuit encodes features of the auxiliary complexity geometry. We have shown how the complexity geometry metric is encoded holographically. It would be interesting to understand if other characteristics of the circuit geometry, for instance the sectional curvature considered in the context of circuits underlying our work in~\cite{Flory:2020eot,Flory:2020dja} and also of broader interest for complexity in quantum-many body systems~\cite{Auzzi:2020idm,Wu:2021pzg,Brown:2021euk}, acquire a natural gravitational interpretation.

It would also be interesting to extend our setup to the lower-dimensional duality between Jackiw–Teitelboim (JT) gravity and random matrix theory.
An interesting feature of this duality is that the partition function of JT gravity cannot be written as a trace over a Hilbert space of $e^{-\beta H}$ due to it being dual to an average over random matrices \cite{Stanford:2017thb,Saad:2019lba}.
Therefore, the expression on the right hand side of \eqref{eq:thermal-two-point function} whose bulk dual we found in AdS$_3$ is not equal to the Fubini-Study distance in the TFD state as it does not have the interpretation of a connected two-point function of the Hamiltonian in the TFD state.
Rather, to determine the averaged Fubini-Study distance one has to compute 
\begin{equation}
   \overline{\langle H^2 \rangle_\beta-\langle H \rangle_\beta^2} = \overline{\frac{\partial_\beta^2 Z(\beta)}{Z(\beta)} - \left(\frac{\partial_\beta Z(\beta)}{Z(\beta)}\right)^2},
\end{equation}
i.e.~first compute the Fubini-Study distance for a single member of the random matrix ensemble and then average.
Finding a JT gravity dual to this averaged Fubini-Study distance would allow for studying quantum corrections at late times.
Such corrections were already found to lead to a plateau in the holographic complexity measures in JT gravity studied in \cite{Iliesiu:2021ari,Alishahiha:2022kzc,Alishahiha:2022exn}.
Ensemble averages of CFTs in two and higher dimensions have recently attracted considerable attention in relation to new Euclidean wormhole contributions to the gravitational path integral (see e.g.~\cite{Cotler:2022rud,Belin:2020hea,Afkhami-Jeddi:2020ezh,Maloney:2020nni}).
Therefore, generalizing our setup to JT gravity will likely provide further clues if and how holographic complexity measures in higher dimensions such as the one studied here saturate at late times.

Finally, the Virasoro algebra that formed a crucial ingredient in our construction also emerges in a variety of discrete systems such as spin chains and their tensor network description~\cite{Koo:1993wz,Milsted:2017csn,Zou:2017zce,Milsted:2018vop,Milsted:2018yur,Zou:2019dnc,Milsted:2018san,Hu:2017rsp}. It would be interesting to connect with complexity studies in these quantum-many body setups, in light of recent advancements \cite{Brown:2021euk,Brown:2021rmz, Brown:2022phc} on computational complexity in discrete systems in the framework of \cite{Nielsen} that we also employed here.

\acknowledgments

We would like to thank Mario Flory for his involvement in the early stages of this project. Moreover, we thank Pawe{\l} Caputa, Shira Chapman, Bartek Czech, Jan de Boer, Juan Hernandez, Mikhael Khramtsov, Rob Myers and Tadashi Takayanagi for useful discussions and correspondence. The work of A.-L.~W. is supported by DFG, grant ER 301/8-1 | ME 5047/2-1. The work M.~G. was supported by Germany's Excellence Strategy through the W\"urzburg‐Dresden Cluster of Excellence on Complexity and Topology in Quantum Matter ‐ ct.qmat (EXC 2147, project‐id 390858490). This research has been supported by FWO-Vlaanderen project G012222N and by Vrije Universiteit Brussel through the Strategic Research Program High-Energy Physics.
This research was supported in part by Perimeter Institute for Theoretical Physics. Research at Perimeter Institute is supported by the Government of Canada through the Department of Innovation, Science and Economic Development and by the Province of Ontario through the Ministry of Research, Innovation and Science.

\appendix

\section{Lessons from SL(2,R) circuits}
\label{sec:SL(2,R)-circuits}
In this appendix we compare our geometrization of the Fubini-Study metric in the holographic bulk spacetime with the construction in \cite{Chagnet:2021uvi} for SL(2,R) circuits.

The authors of \cite{Chagnet:2021uvi} considered circuits implementing global conformal transformations in $d\geq 2$. For a more direct comparison with our work, we discuss only CFTs in two dimensions. In this setup, the global conformal transformations form the symmetry group SL(2,R), and diffeomorphims may be parameterized in terms of three parameters $\gamma_R(\tau),\zeta(\tau),\zeta^*(\tau)$, 
\begin{equation}
    F(\tau, x^+)=-i \log \left(\frac{i e^{i\left(x^++\gamma_R(\tau)\right)}-\zeta(\tau)}{i+e^{i\left(x^++\gamma_R(\tau)\right)} \zeta^*(\tau)}\right),
    \label{eq:diffeo_tau}
\end{equation}
where $x^\pm$ are lightcone coordinates on the cylinder. The circuit corresponding to \eqref{eq:diffeo_tau} is given by
\begin{equation}
    U(\tau) \ket{h}\equiv e^{i \zeta(\tau) L_{-1}} e^{i \gamma(\tau) L_0} e^{i \zeta_1(\tau) L_1}\ket{h},
\end{equation}
where $\gamma(\tau)=\gamma_R(\tau)-i \log(1-|\zeta|^2)$ and $\zeta_1(\tau)=\zeta^{*}(\tau)e^{i \gamma_R(\tau)}$.
Note that the transformation \eqref{eq:diffeo_tau} is a symmetry of the CFT in the vacuum state. Therefore, an appropriate reference state for such circuits is $\ket{h}$ with $h>0$.
The Fubini-Study distance for \eqref{eq:diffeo_tau} is given by
\begin{equation}
    	\mathrm{d} s_\text{FS}^{2}=2 h\frac{\mathrm{d} \zeta \mathrm{d} \zeta^{*}}{\left(1-|\zeta|^{2}\right)^{2}}.
    \label{eq:FS_SL2R}
\end{equation}

In the gravity theory, the reference state $\ket{h}$ in the CFT circuit corresponds to a conical defect geometry,
\begin{equation}
		 ds_{\mathrm{AdS}}^2=d\rho^2-\cosh{\rho}^2dt^2+\sinh^2(\rho)d\phi^2,
		 \label{eq:metric_global_AdS}
\end{equation}
where $\phi$ is $\frac{2\pi}{n}$-periodic and $n\in \mathbb{N}$. The geometry can be interpreted as empty AdS with a particle of mass $m$ located at the center $\rho=0, t=0, \phi=0$. As it moves along a timelike trajectory, the massive particle cuts out a wedge related to its mass, $m=-1 / 8 G n^2$ \cite{Chen:2018vkw}.

The authors of \cite{Chagnet:2021uvi} then observed that the symplectic geometry associated to the circuit generated by the transformations \eqref{eq:diffeo_tau} is equivalent to that of timelike geodesics in AdS. This equivalence indicates that geometric features of the phase space such as the Fubini-Study distance \eqref{eq:FS_SL2R} may be written in terms of timelike geodesics. The timelike geodesics corresponding to the circuit arising from \eqref{eq:diffeo_tau} may be obtained by considering the empty AdS geodesic $(\rho(t),t,\phi(t))=(0,t,0)$ and boosting it by \eqref{eq:diffeo_tau}, which yields a sequence of geodesics in a single geometry. It is convenient to rewrite the geodesics in terms of embedding-space coordinates,
\begin{equation}
\begin{aligned}
&T_1=\cosh (\rho(t)) \cos (t), \\
&T_2=\cosh (\rho(t)) \sin (t), \\
&X_1=\sinh (\rho(t)) \cos (\phi(t)), \\
&X_2=\sinh (\rho(t)) \sin (\phi(t))
\end{aligned}
\end{equation}
with metric $d s^2=-d T_1^2-d T_2^2+d X_1^2+d X_2^2$. If we apply only a left-moving transformation \eqref{eq:diffeo_tau}, the geodesics read
\begin{equation}
    \begin{aligned}
        \phi(t)&=\arctan(i+\frac{2 i\zeta^*}{e^{2 i t} \zeta-\zeta^*}),\\
        \rho(t)&=\mathrm{arcsech}(\sqrt{1-|\zeta|^2}).
    \end{aligned}
\end{equation}
By $X^{\mu}$ we now denote the vector $ X^{\mu}=\{X_1,X_2\}$. 
The Fubini-Study distance is then given in terms of the minimal and maximal distance between two such geodesics~\cite{Chagnet:2021uvi}, 
\begin{equation}
 \mathrm{d} s_{\mathrm{FS}}^2=\frac{h}{2 }\left(\delta X_{\mathrm{perp}, \min }^2+\delta X_{\mathrm{perp}, \max }^2\right).
 \label{eq:dual_FS_timelike}
\end{equation}

Let us now highlight some important differences to our construction of a dual to the Fubini-Study metric. 
First of all, the result \eqref{eq:dual_FS_timelike} relies on the identification of the configuration space $(\zeta,\zeta^*,\gamma_R)$ associated to the circuit and the phase space of timelike geodesics in AdS. It is therefore not straightforward to generalize \eqref{eq:dual_FS_timelike} to geometries other than the conical defect and to transformations beyond SL(2,R). A general conformal transformation in two dimensions, which is the focus of our work, is parameterized by an infinite number of parameters and therefore can no longer be identified with the phase space of a massive particle. Furthermore, in \cite{Chagnet:2021uvi}, there is no dual to the circuit itself in the bulk geometry. Instead, a state in the CFT corresponds to a geodesic along its full trajectory and multiple boosted geodesics representing the states in the circuit are considered within the same geometry. On the other hand, in our construction, the state is defined on a constant time slice in the CFT and extends into the bulk to the full Wheeler-De Witt patch. The evolution of the state according to the circuit is encoded in the time evolution of the geometry. In particular, physical time has no significance in \eqref{eq:dual_FS_timelike} and is eliminated by the extremization procedure. 

Finding a similar realization of the Fubini-Study distance in terms of timelike geodesics in our setup is not straightforward as the timelike geodesics considered in \cite{Chagnet:2021uvi} differ in another important aspect from our construction: The $U(1)$ transformation   $x^+\rightarrow x^+ +\gamma_R(\tau)$ constitutes a global symmetry and leaves the the timelike geodesic invariant. The $U(1)$ transformation acts on each constant time slice by shifting it by a value $\gamma_R(\tau)$ that is constant with respect to the physical time $t$. In our construction, the identification $\tau=t$ implies that a similar $U(1)$ transformation $x^+\rightarrow x^+ +\alpha(t)$ acts differently on each constant time slices as the shift $\alpha(t)$ now depends on physical time. Timelike geodesics in this case start exhibiting $\alpha(t)$-dependence. Since we know that $\alpha(t)$ does not enter the Fubini-Study distance, it is therefore natural to consider a geometric object that we already know to share this property instead. This is the length of spacelike geodesics on a constant time slice, which is the main ingredient of our cost measure, as discussed in \secref{sec:bulk-dual-Fubini-Study}.

\section{The switchback effect in delocalized and black string shockwaves}
\label{sec:further-shockwaves}
In order to determine how universal the switchback effects found in \secref{sec:switchback} are, in this appendix we study the geometric expression $F_\text{bulk}$ from \eqref{eq:definition-F_bulk} in further shockwave geometries than the ones considered in \secref{sec:switchback}.
We consider shockwaves in a black string geometry, i.e.~a black hole with non-compact horizon, as well as delocalized shockwaves where the matter is concentrated uniformly in the angular direction.
We find that in both cases the switchback effect is reproduced qualitatively in the sense that $F_\text{bulk}$ increases with increasing $t_W$ before saturating to the variance of the Hamiltonian in the TFD state.
However as discussed below, from a quantitative perspective the match to the expectations from the infection models from \cite{Stanford:2014jda,Susskind:2018pmk} is not quite as good as for the localized shockwaves in the BTZ black hole geometry considered in \secref{sec:switchback}.

\subsection{Delocalized shockwaves}
\begin{figure}
  \centering
  \includegraphics[width=.3\textwidth]{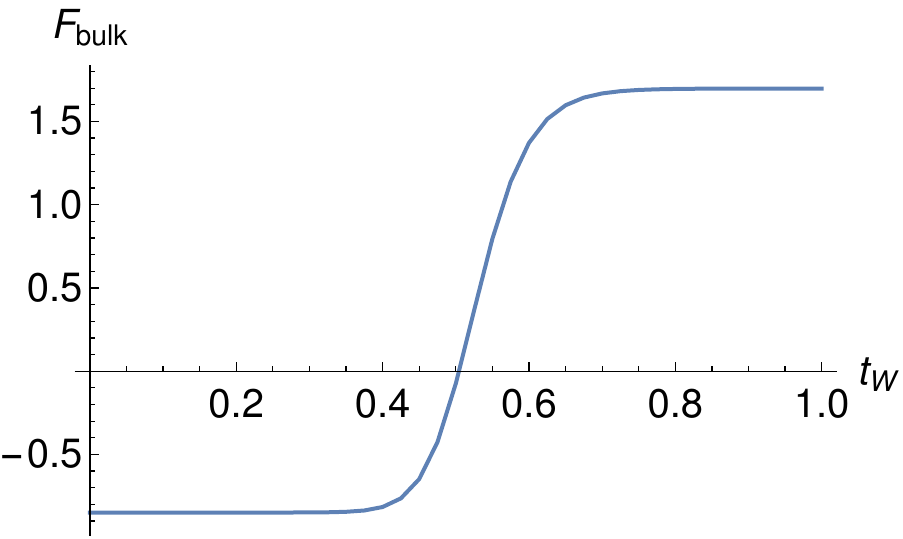}
  \quad
  \includegraphics[width=.3\textwidth]{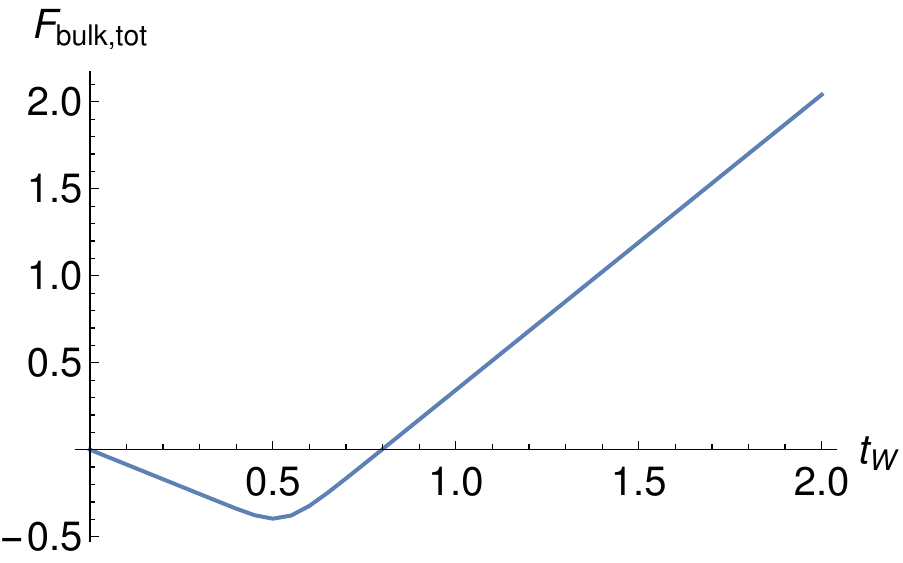}
  \caption{Left: $F_\text{bulk}$ plotted over $t_W$ and right: the corresponding total cost $F_\text{bulk,tot}$ (the $t_W$ integral of $F_\text{bulk}$) in a BTZ geometry perturbed by a single delocalized shockwave with strength $\alpha$ from \eqref{eq:alpha-parameter} and $t_*=0.5$, $\beta=1/4$.
  }
  \label{fig:shockwave-Fubini-Study}
\end{figure}

We will start with simple shockwave geometries where the shockwave is delocalized in the space direction (but still localized at the horizon) used for instance in \cite{Stanford:2014jda}.
In this case, the solution of Einstein's equation is given by \eqref{eq:shockwave-metric} with a displacement function $h(\phi)=\alpha$ that is constant along the spatial direction.
The geodesic length from the left boundary to the point $(u=0,v=v_s+\alpha,\phi=\phi_s)$ on the left side of the shockwave is given by
\begin{equation}
  \ell_1 = \log\left[\frac{e^{r_H t_1}(v_s+\alpha) + \cosh(r_H(\phi_1-\phi_s))}{\sqrt{\epsilon_\text{UV}}}\right],
\end{equation}
and the length from the point $(u=0,v=v_s,\phi=\phi_s)$ on the right side of the shockwave to the right boundary by
\begin{equation}
  \ell_2 = \log\left[\frac{-e^{-r_H t_2}v_s + \cosh(r_H(\phi_2-\phi_s))}{\sqrt{\epsilon_\text{UV}}}\right].
\end{equation}
The total geodesic length is given by
\begin{equation}
  \ell = \log\left[\frac{1}{\epsilon_\text{UV}}\left(\sqrt{\frac{1}{2}\cosh(r_H(\phi_1-\phi_2))+\frac{1}{2}\cosh(r_H(t_1-t_2))} + \frac{\alpha}{2}e^{r_H(t_1-t_2)/2}\right)^2\right].
\end{equation}
Again, we find a time delay in $t_W$ before the Fubini-Study distance saturates (see \figref{fig:shockwave-Fubini-Study}) and thus the total Fubini-Study cost \eqref{eq:Fubini-Study cost} increases linearly with a time delay.
These features qualitatively reproduce the switchback effect.
However, in this case $F_\text{bulk}$ is negative in some parameter regimes which is not sensible for a cost function for computational complexity and hence $F_\text{bulk}$ can no longer be interpreted as the gravity analog of the number of infected qubits in the model of \cite{Stanford:2014jda,Susskind:2018pmk}.
As the delocalized shockwaves are more akin to a delocalized perturbation $W$ acting on the whole system instead of the localized perturbation acting on a single qubit considered in \cite{Stanford:2014jda,Susskind:2018pmk}, this match in qualitative but not quantitative terms is not too surprising.

\subsection{Localized shockwaves with non-compact horizon}
Another class of shockwave geometries are those with non-compact horizon and localized matter concentrations studied in \cite{Roberts:2014isa}.
In this case, a localized shockwave corresponds to a metric \eqref{eq:shockwave-metric} with displacement
\begin{equation}
  v \to v + h(\phi) \quad \text{with} \quad h(\phi) = \alpha\, e^{-|\phi - \hat\phi|r_H},
\end{equation}
where $\hat\phi$ is a constant parameter and $-\infty < \phi < \infty$.
This is obtained by solving Einstein's equations with an energy concentration localized at the $u=0$ horizon and at $\phi = \hat \phi$ \cite{Roberts:2014ifa},
\begin{equation}
  T_{uu} = \frac{\alpha}{2\pi G_N} \delta(u)\delta(\phi - \hat \phi) \quad \text{and} \quad T_{\mu\nu} = 0, \quad (\mu,\nu) \neq (u,u).
\end{equation}

The geodesic lengths are given by
\begin{equation}
  \begin{aligned}
    \ell_1 &= \log\left[\frac{e^{r_H t_1}(v_s+h(\phi_s)) + \cosh(r_H(\phi_1-\phi_s))}{\sqrt{\epsilon_\text{UV}}}\right],\\
    \ell_2 &= \log\left[\frac{-e^{-r_H t_2}v_s + \cosh(r_H(\phi_2-\phi_s))}{\sqrt{\epsilon_\text{UV}}}\right].
  \end{aligned}
\end{equation}
and
\begin{equation}
  \ell = \log\left[\cosh(r_H(\bar t \pm \Delta\phi))\left(\cosh(r_H(\bar t \mp \Delta\phi)) + \alpha e^{r_H(\pm\hat\phi \mp \bar\phi + \Delta t)}\right)\right],
\end{equation}
with $\pm = \mathrm{sgn}\left(\bar\phi - \hat\phi - \frac{1}{2r_H}\log\left(\frac{\cosh(r_H(\bar t - \Delta\phi))}{\cosh(r_H(\bar t + \Delta\phi))}\right)\right)$ where $\bar\phi = (\phi_1+\phi_2)/2$, $\Delta\phi=(\phi_1-\phi_2)/2$, $\bar t = (t_1+t_2)/2$, $\Delta t=(t_1-t_2)/2$.
The results for $F_\text{bulk}$ from \eqref{eq:bulk-dual-FS-distance} are qualitatively similar to the case of a localized shockwave with compact horizon studied in \secref{sec:switchback}.
Again, we find an increase of $F_\text{FS}$ followed by a saturation if $t_W$ and with it the shockwave strength $\alpha$ is varied while $t=0$ is kept constant.
Unlike for the compact horizon case, however, here we find that $F_\text{bulk}$ does not asymptote to zero for $t_W \to 0$.
In the infection model of \cite{Stanford:2014jda,Susskind:2018pmk}, this may be interpreted as a perturbation $W$ which is so large that already at $t_W=0$ a sizeable fraction of the system is infected.

\bibliographystyle{utphys}
\bibliography{bibliography}

\end{document}